\documentclass[review]{elsarticle}

\usepackage{lineno,hyperref}

\usepackage{array}
\usepackage{multirow}
\usepackage{tabularx}
\usepackage{pdflscape}
\usepackage{xcolor}
\usepackage{xurl}
\hypersetup{breaklinks=true} 

\graphicspath{ {./figures/} }

\biboptions{authoryear}

\begin{document}

\begin{frontmatter}

\title{Degrees of displacement: The impact of household PV battery prosumage on utility generation and storage}

\author[Curtin,ETHub]{Kelvin Say\corref{correspondingauthor}}
\cortext[correspondingauthor]{Corresponding author}
\ead{kelvin.say@unimelb.edu.au}

\author[DIW,ETHub]{Wolf-Peter Schill}
\ead{wschill@diw.de}

\author[Curtin]{Michele John}
\ead{m.rosano@curtin.edu.au}

\address[Curtin]{Sustainable Engineering Group, Curtin University, Kent Street, Bentley 6102, Australia}
\address[ETHub]{Energy Transition Hub. University of Melbourne, Grattan Street, Parkville 3010, Australia.}
\address[DIW]{German Institute for Economic Research (DIW Berlin), Mohrenstra{\ss}e 58, 10117 Berlin, Germany}

\begin{abstract}
Reductions in the cost of PV and batteries encourage households to invest in PV battery prosumage. We explore the implications for the rest of the power sector by applying two open-source techno-economic models to scenarios in Western Australia for the year 2030. Household PV capacity generally substitutes utility PV, but slightly less so as additional household batteries are installed. Wind power is less affected, especially in scenarios with higher shares of renewables. With household batteries operating to maximise self-consumption, utility battery capacities are hardly substituted. Wholesale prices to supply households, including those not engaging in prosumage, slightly decrease, while prices for other consumers slightly increase. We conclude that the growth of prosumage has implications on the various elements of the power sector and should be more thoroughly considered by investors, regulators, and power sector planners.
\end{abstract}



\end{frontmatter}

\pagebreak

\section{Introduction}

To mitigate the effects of climate change it is necessary to take advantage of renewable energy sources and decarbonise energy use \citep{de_coninck_strengthening_2018}. Continued investments in research and development as well as the massive deployment of renewable energy technologies has reduced the Levelised Costs of Energy of PV and wind power in many regions to or below those of conventional fossil fuel generation \citep{bnef_new_2019, irena_renewable_2019, ren21_renewables_2019}. These ongoing cost reductions have not only changed how utilities generate their electricity, but have also opened new opportunities for electricity customers \citep{stephan_limiting_2016}. In combination with favourable regulatory settings, it has become increasingly attractive for households to install their own PV systems in many countries. This not only allows households to reduce their electricity bills but also decarbonises their energy consumption \citep{agnew_effect_2015}.

A similar transition is occurring in the lithium-ion battery sector with global manufacturing capacity expanding to supply the expected growth in the battery electric vehicle market \citep{curry_lithium-ion_2017}. This is driving significant cost reductions, which are expected to continue decreasing at a lower rate over the next 20 years \citep{irena_electricity_2017, lombrana_lithium_2019, schmidt_future_2017}. These battery cost reductions have led to a growing number of utility and domestic-scale battery installations in electricity markets worldwide \citep{eia_us_2019, european_union_summary_2018, irena_innovation_2019}. By storing excess PV generation for later use, PV-battery systems enable households to increase their overall share of self-generation. This concept, referred to as prosumage \citep{bustos_evolution_2019, green_prosumage_2017, schill_prosumage_2017}, can also significantly reshape grid consumption and retailer revenues \citep{say_power_2019}.

This paper aims to quantitatively explore the influence that residential PV systems with or without batteries could have on the power sector, in particular on utility-scale generation and storage technologies. We do so by applying two soft-coupled open-source models to 2030 scenarios in the South West Interconnected System (SWIS) located in Western Australia. This serves as a particularly suitable case study, as household PV penetration rates here are amongst the highest in the world \citep{apvi_pv_2019} and household PV-battery installations are also beginning to rise \citep{aec_solar_2019}. As it is an island network,\footnote{Island networks face more challenges in matching variable renewable energy supply with demand compared to interconnected networks, as they lack the ability of balancing over larger regions.} the power sector effects from prosumage become evident earlier than in larger and interconnected networks. Firstly, we use a techno-economic simulation model of household prosumagers, \textit{Electroscape} \citep{say_power_2019}, in which a set of heterogeneous households are driven by economic self-interest to invest in additional PV and battery capacity while retail price conditions change under different Feed-in Tariff (FiT) values. By using these households as representatives for the segment of customers investing in prosumage, we quantify the changes to the ‘residual network demand’, also known as ‘operational demand’ \citep{aemo_operational_2019} or ‘net load’ \citep{odwyer_using_2015}. This serves as an input for a dispatch and investment model, which determines cost-minimal utility-scale generation and storage capacity while meeting different exogenous renewable energy targets.

This paper is structured as follows: Section \ref{sec: background and lit} presents a background and literature review. Section \ref{sec: methods} introduces the underlying methodology and modelling framework. Section \ref{sec: input data and case study} describes the input data. Section \ref{sec: results} presents the results and wider implications on the power sector. Section \ref{sec: key assumptions} discusses some key assumptions of the study and their qualitative impacts on results. Section \ref{sec: conclusions} discusses and concludes with policy implications and avenues for future research.

\section{Background and literature review\label{sec: background and lit}}

Australia currently leads the world in household PV adoption and the substantial existing PV capacity situated behind-the-meter raises the potential for an accelerated PV-battery transition, especially as the financial benefits from PV-battery systems begin to outweigh PV-only systems. While PV-only and PV-battery systems are both considered as behind-the-meter Distributed Energy Resources (DER), their grid utilisation and economic drivers considerably differ and require more detailed analysis. The term prosumage, in this paper, covers both household PV-only and PV-battery adoption.\footnote{We slightly expand the narrower definition of prosumage used in \cite{schill_prosumage_2017} to avoid lengthy verbal differentiations when describing results for PV-only and PV-battery cases.}

\subsection{Australia as a front-runner in household PV and battery adoption\label{subsec: intro Australia}}

High levels of solar insolation, relatively high volumetric retail tariffs \citep{aemc_2019_2019}, and residential FiT policies\footnote{The FiT is only applied to the amount of excess solar PV energy generated after subtracting the customer’s underlying electricity demand. FiTs in Australia are typically valued well below volumetric retail tariffs \citep{aec_solar_2019}. FiT payments are funded by electricity retailers and revised annually (as opposed to fixed-term contracts) \citep{poruschi_revisiting_2018}.} have resulted in over 2 million Australian households (or 20\% of all free-standing households) installing solar PV systems \citep{apvi_pv_2019}. As of the end of 2018, combined household PV capacity ($7~GW_P$) accounted for 62\% of the nation’s installed solar PV capacity \citep{apvi_pv_2019}. As discounted payback periods approach or fall under 5 years in most Australian capital cities \citep{aec_solar_2019}, household PV installations are expected to continue rising \citep{graham_projections_2018}. These household PV systems, that reside behind the meter and are neither centrally monitored nor controlled, are no longer insignificant and have begun to reshape residual network demand and system operation \citep{aemo_integrating_2019, aemo_technical_2019, aemo_quarterly_2019}. The inability to control DER systems behind-the-meter \citep{aemo_integrating_2019} also effectively grants household generation the highest dispatch priority on the network, followed by zero-marginal cost and non-dispatchable utility PV and wind, then conventional baseload and peaking generation. With regards to battery energy storage, both utility \citep{aurecon_hornsdale_2018, irena_innovation_2019, neoen_hornsdale_2017} and household \citep{sunwiz_australian_2018} installations have begun to rise in Australia, as costs decrease with increased global production \citep{schmidt_future_2017}.

The SWIS network in Western Australia has a typical peak demand of $4.4~GW$ and $18~TWh$ of annual operational consumption. Residential PV capacity continues to rise and at times in 2019 has been able to supply approximately 45\% of the underlying network demand \citep{aemo_quarterly_2019}. With a household PV penetration rate above 27\%, it has one of the highest rates in the world \citep{aec_solar_2019, apvi_pv_2019}. Its cumulative installed household PV capacity at the end of 2019 was $800~MW_P$, and is forecasted to reach $2~GW_P$ in 2030 \citep{aemo_wem_2019-2}. Since 2014, household battery adoption has increased year-on-year and is expected to continue growing at an accelerated rate \citep{aec_solar_2019, aemo_wem_2019-2}. The SWIS network is islanded and lacks the means to balance load with neighbouring networks. As installed capacities grow, prosumage households in the future could have considerable influence on the appropriate mix of utility-scale generation and storage technologies.

\subsection{Studies of PV-only and PV-battery households\label{subsec: intro prosumage}}

With household PV-only systems, the timing of all self-generation is tied to the sun without the ability to store and buffer energy. This means that changes in residual grid consumption begin to coincide with other households, thus driving observable grid demand patterns such as the 'duck curve' \citep{denholm_overgeneration_2015, maticka_swis_2019}. The use of energy storage, changes the level of coincidence by making excess PV generation available for later use and increasing the system's sensitivity to the type of economic incentives and differences in household demand. This not only changes the overall residual grid consumption, but also the effectiveness of existing FiT policies to guide PV battery adoption \citep{gunther_prosumage_2019, say_power_2019}.

By adopting battery storage, households become technically capable of providing further services to the rest of the power sector. Since the supply and demand of energy must always be in balance, spare household battery capacity could be used as a form of dispatchable load or generation to provide quantifiable system benefits \citep{lawder_battery_2014, von_appen_strategic_2018}. However the use of time-invariant volumetric residential tariffs remains common in many regions, including Australia \citep{simshauser_inequity_2016, nelson_electricity_2018}, Europe \citep{jargstorf_assessing_2015}, UK \citep{nicolson_are_2017}, and China \citep{lin_designation_2012}. The time-invariant nature of these tariffs does not give households an incentive to consider wholesale market price signals when operating their PV battery systems, thus leaving increased PV self-consumption as the largest financial incentive for households to invest in PV battery capacity \citep{rocky_mountain_institute_economics_2015}.

At the household scale, techno-economic models and electricity bill savings are commonly used to determine the appropriate sizing of household PV-battery systems. Using project finance metrics, such as Net Present Value (NPV) \citep{say_power_2019, schopfer_economic_2018, talent_optimal_2018}, Internal Rate of Return \citep{parra_effect_2016} and Discounted Payback Periods \citep{akter_comprehensive_2017}, optimal system capacities can be calculated. At the utility scale, techno-economic models are commonly used for long term energy planning and renewable energy integration. Using numerical optimisation, many different objectives can be evaluated, such as least-cost utility-scale renewable energy portfolios \citep{jeppesen_least_2016}, coordinating renewable generation, network and storage expansion \citep{haller_bridging_2012}, through to establishing optimal utility-scale energy storage capacities \citep{schill_long-run_2018}. The objectives of these household and utility models differ, with households aiming to reduce electricity bills, and techno-economic models aiming to reduce the overall cost supplying energy. Hence, modelling frameworks that combine these perspectives are rare in the literature. 

\subsection{Main contributions of this paper\label{subsec: intro contributions}}

This paper addresses this gap by using two open-source models to link household PV battery investment decisions and optimal utility-scale generation and storage decisions from a social planner perspective. A counterfactual comparison is used to provide quantitative insights into the range of utility-scale system impacts from household prosumage, including generation and storage capacities, their dispatch and wholesale price impacts.

The Western Australian (WA) case study investigated in this paper allows the derivation of relatively undistorted insights into the effects of prosumage, as WA has a liberalised electricity market and islanded network. Given the real-world conditions that are currently driving Western Australia’s significant household PV and growing battery adoption, these scenario analyses provide a front-runner case of what other markets could expect in the future. Moreover, the development and provision of the two open-source models also contribute to the literature by providing transparency and reproducibility for subsequent research.

\section{Methods\label{sec: methods}}

\subsection{General setup}

\begin{figure}[ht]
\centering{} \includegraphics[width=1\textwidth]{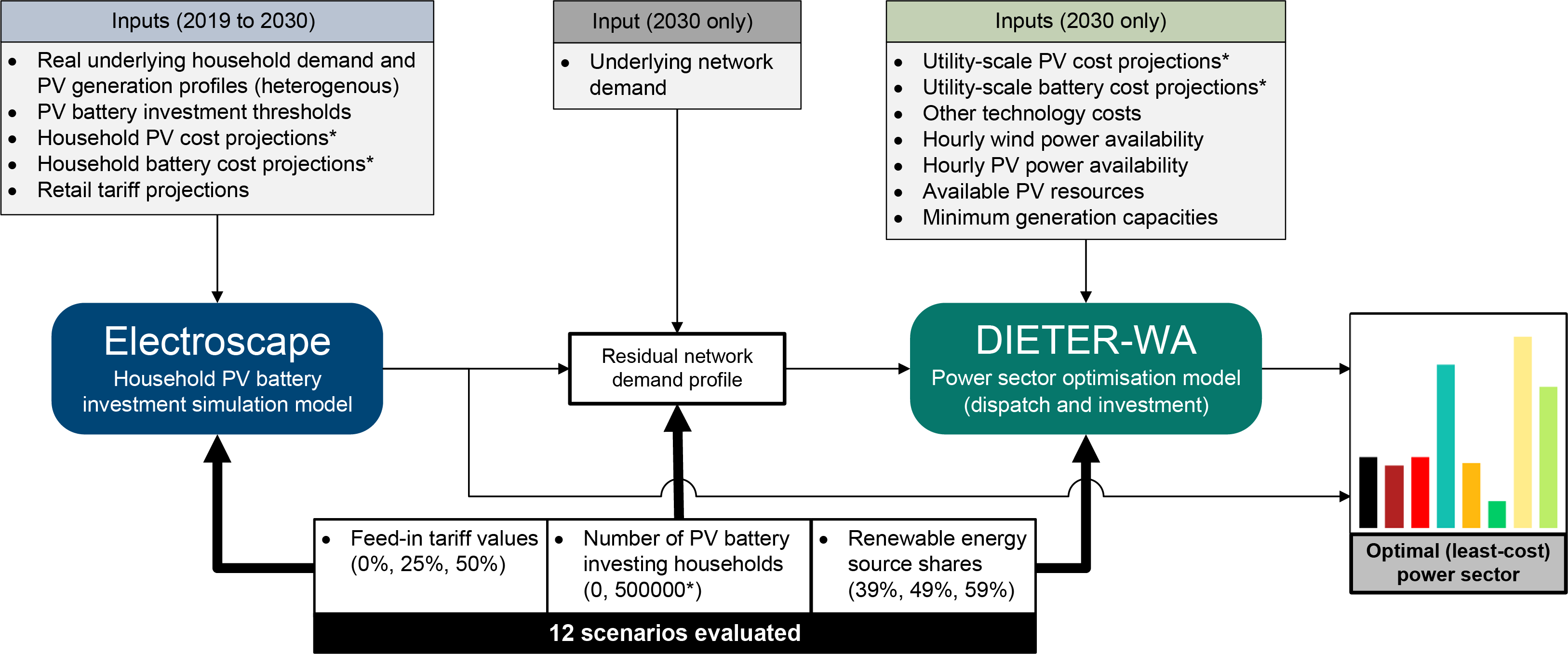}\\
\protect\caption{\label{fig: methodology}Developed methodology that integrates household prosumage and utility-scale investment and dispatch decisions. Starred parameters are varied in sensitivity analyses.}
\end{figure}

We soft-link two open-source techno-economic models to represent the differing objectives between PV battery investing prosumage households and central planner investments in utility-scale generation and storage capacity (Figure \ref{fig: methodology}). The first model \textit{Electroscape} reflects the financial objectives of prosumage households as retail conditions change over time. These households consider investing annually in PV and battery systems, given exogenous assumptions on retail price conditions, installed system costs and three FiT scenarios (0\%, 25\%, 50\% of volumetric usage charges) between 2019 and 2030. The second model \textit{DIETER-WA} adopts a central planner perspective for the overall power sector, i.e.,~it determines least-cost utility-scale investment and dispatch decisions over a range of locally available technologies in 2030. \textit{DIETER-WA} uses outcomes from \textit{Electroscape} and its three FiT scenarios to also assess the additional impact of varying the Renewable Energy Source (RES) share between 39\%, 49\%, and 59\% of gross electricity demand, where 49\% is a linear interpolation between Australia’s 2020 renewable energy target of 23.5\% \citep{commonwealth_of_australia_renewable_2019} and an assumed 100\% target for 2050.\footnote{Based on Australia’s commitment to the COP21 Paris Agreement \citep{commonwealth_of_australia_paris_2016}.} By comparing each of these results against reference counterfactual scenarios (without household PV battery investments) the effects of household prosumage on the overall system are quantified and separated.

Both models are soft-linked through an hourly time series of the residual network demand profile nad household prosumage investments. \textit{Electroscape} is solved annually for each of the FiT scenarios between 2019 and 2030 to determine the resulting net grid utilisation profile in the year 2030 for each individual prosumage household. These profiles are normalised and scaled to an assumed number of 500,000 prosumage households, which builds on the independent system operator’s PV installation estimates for 2030 \citep{aemo_wem_2019-2}. By subtracting the net grid utilisation changes from prosumage households from actual SWIS network demand data \citep{aemo_wem_2019}, we determine the overall impact of household PV and battery investments on the residual network demand. To isolate the effect of prosumage household investments, all other customers of the SWIS electricity network are assumed to consume the same amount of energy each year without investing in self-generation or energy efficiency.

\subsection{Household PV battery investment modelling (\textit{Electroscape})}

To determine PV battery investment decisions for residential households, \textit{Electroscape} \citep{say_power_2019} uses the time-series of underlying household demand and insolation profiles with projections of retail tariffs, FiTs and PV battery installation costs and takes into consideration previous PV battery investments. By evaluating households individually and with real-world energy meter data, the model avoids biases that can be introduced when using aggregated or synthesised data \citep{quoilin_quantifying_2016, schopfer_economic_2018}. This model was first introduced in \cite{say_coming_2018} and subsequently used to evaluate the relationship between household PV battery investments and future electricity retailer revenues \citep{say_power_2019}. The model is implemented in R with its source code and data available under permissive open-source license \citep{Say_Electroscape}.

Assuming that prosumage households are economically rational, the model simulates the installation of each and every PV and battery combination (using a step size of $0.5~kW_P$ and $1~kWh$ respectively) on a household's underlying demand to determine the net grid utilisation and annual electricity bills. By comparing against a ‘no installation’ case, the expected savings in electricity bills are calculated and form the basis of a discounted cashflow for each and every combination. This model assumes a fixed investment horizon and uses Net Present Value (NPV) to compare each PV battery configuration as competing investment opportunities. The configuration with the highest NPV becomes a prime candidate for installation, pending a real options valuation \citep{reuter_renewable_2012} based on Discounted Payback Periods (DPPs),\footnote{The DPP is used to publicly track potential PV system financial performance in Australia \citep{aec_solar_2019} which is mirrored in this model.} the model determines if sufficient financial returns can be realised to warrant making an actual investment. If an investment is made, the underlying household demand is updated, and subsequent PV battery investments must now consider these newly installed systems. Repeating this process annually allows \textit{Electroscape} to simulate sequential and lumpy PV battery investment behaviour that reacts to changing retail conditions, namely retail tariffs, FiTs and PV battery installation costs. The model provides the means to model PV battery investment choices catered to the energy use and solar resources of individual households. By applying \textit{Electroscape} across a set of real and heterogeneous household data, the normalised results are able to provide an approximate of the net grid utilisation from an average prosumage household.

\subsection{Power sector dispatch and investment modelling (\textit{DIETER-WA})}

To investigate the power sector effects of increased prosumage, we devise the open-source model \textit{DIETER-WA}. It represents a simplified and adjusted version of the dispatch and investment model \textit{DIETER}, which has been first introduced by \cite{zerrahn_long-run_2017} and \cite{schill_long-run_2018}. The model has a long-run equilibrium perspective and minimises the total cost of utility-scale electricity generation for all subsequent hours of a whole year. Its results may be interpreted from a central planner perspective, or as an outcome of a frictionless market with perfect competition. The model assumes perfect foresight and is solved for all consecutive hours of an entire year. It is implemented in the General Algebraic Modelling System (GAMS). Source code and input data are available under a permissive license \citep{Schill_DIETER-WA}.

The model’s objective function covers operational costs which consist of fuel and other variable costs, as well as annualised investment costs of all utility-scale generation and storage technologies. An energy balance ensures that electricity supply satisfies demand in each hour. Generation technologies comprise both dispatchable thermal and variable renewable generators. The model is also capable of representing various energy storage technologies and their respective intertemporal restrictions. In the model version used here, we ensure that a specified share of yearly gross electricity demand is met by renewable energy sources, including household PV installations.

Model inputs comprise specific fixed and variable costs of all technologies, hourly renewables availability factors, as well as the residual network demand profile, which considers the net grid utilisation profiles of prosumage households determined by \textit{Electroscape}. Prosumage PV and battery investments also enter as exogenous inputs. Endogenous variables include investments in utility-scale generation and storage technologies and their hourly use. Further model outputs comprise the total cost of providing electricity and the shadow prices of the energy balance equation, which we interpret as wholesale prices.

\section{Input data\label{sec: input data and case study}}

\subsection{Input data for \textit{Electroscape}\label{subsec: input electroscape}}

The main input parameters and data used in \textit{Electroscape} are summarised in Table \ref{tab: inputs electroscape}.

\begin{table}[ht]
\tiny
\begin{center}
\begin{tabular}{ l l l l }
\hline
\textbf{Input parameter}	&	\textbf{Unit}	&	\textbf{Values}	&	\textbf{Source}	\\
\hline \\
Scenario forecast period	&	years	&	12	&	Own assumption	\\
Simulation time step	&	minutes	&	30	&	Own assumption	\\
Financial horizon	&	years	&	10	&	Own assumption	\\
DPP evaluation criteria	&	years	&	5	&	Own assumption	\\
Initial PV evaluation range	&	$kW_{P}$	&	0-10	&	Own assumption	\\
Initial battery evaluation range	&	$kWh$	&	0-18	&	Own assumption	\\
Battery energy-to-power ratio	&	ratio	&	2.5	&	Own assumption	\\
Initial FiT rebate	&	$AUD/kWh$	&	0-14.5	&	Own assumption	\\
Initial volumetric usage charges	&	$AUD/kWh$	&	0.29	&	\cite{infinite_energy_synergy_2019}  \\
Yearly change in tariff charges/rebates	&	\%	&	4	&	\cite{abs_64010_2020}	\\
FiT eligibility limit	&	$kW_{P}$	&	5	&	\cite{synergy_am_2019}	\\
Yearly discount rate	&	\%	&	5	&	\cite{rba_indicator_2019}	\\
Initial installed PV system cost (residential)	&	$AUD/kW_{P}$	&	1292	&	\cite{solar_choice_residential_2019}	\\
Initial installed residential battery system cost	&	$AUD/kWh$	&	1172	&	\cite{solar_choice_battery_2019}	\\
PV cost reduction curves	&	$AUD/kW_{P}$	&	Time series	&	\cite{graham_gencost_2018}	\\
Battery cost reduction curves	&	$AUD/kWh$	&	Time series	&	\citep{schmidt_future_2017}	\\
Number of unique household profiles	&	household	&	261	&	\citep{ausgrid_solar_2018}	\\
Underlying household demand profile (per household) 	&	$Wh$	&	Time series	&	\citep{ausgrid_solar_2018}	\\
Household available insolation profile (per household)	&	$Wh$	&	Time series	&	\citep{ausgrid_solar_2018}   \\	
\hline
\end{tabular}
\caption{Input parameters and data used in \textit{Electroscape}\label{tab: inputs electroscape}}
\end{center}
\end{table}

One year of real utility energy meter measurements of half-hourly resolution ‘underlying household demand’ and ‘insolation’ profiles are used to establish an average representative prosumage household within the SWIS network. This data was collected from 300 households in Sydney, Australia between $1^{st}$ July 2012 and $31^{st}$ June 2013 \citep{ausgrid_solar_2018, ratnam_residential_2017} and has been used in other Australian electricity market studies \citep{ellabban_integrated_2019, babacan_distributed_2017, young_potential_2019, konstantinou_viability_2016}. Due to similar latitudes and climate conditions, the average annual consumption and PV generation profiles in Sydney remain consistent with those of Perth, Australia (which is the primary source of residential demand in the SWIS network). After removing households with missing time series data, 261 households remain for analysis. Data from Sydney households was necessary as strict privacy laws prevent SWIS household data from being publicly available.

The battery model is based on lithium-ion residential systems designed for PV applications, similar to those sold by Tesla,\footnote{\url{https://www.tesla.com/en_AU/powerwall}} LG Chem,\footnote{\url{https://www.lgenergy.com.au/products/battery}} and Sonnen,\footnote{\url{https://sonnen.com.au/sonnenbatterie/}} with a round-trip efficiency of 92\% and 70\% storage capacity remaining at the end of a 10-year operational lifespan. A fixed energy-to-power ratio of $2.5$ is used based on the average of these residential battery systems. Battery systems cost reduction curves are derived from \cite{schmidt_future_2017} and have been scaled with a factor of 0.73  to fit local price conditions \citep{solar_choice_battery_2019}.

The PV generation model assumes a 25-year operational lifespan with 80\% generation capacity remaining. We assume a financial investment horizon of 10 years, reflecting expectations that homeowners typically require profitability before moving to another residence.\footnote{$10.5$~years is the typical duration that a home is owned before being sold \citep{corelogic_corelogic_2015}.} We further assume that households extend their home mortgage to access financial capital and a discount rate of 5\% is used, consistent with the average standard variable home mortgage interest rate over the last 5 years \citep{rba_indicator_2019}. PV cost reduction curves are derived from \cite{graham_gencost_2018} and have been scaled with a factor of $0.78$ to fit local price conditions \citep{solar_choice_residential_2019}.

Corresponding to 2019-20 SWIS retail tariffs, volumetric usage charges begin at $0.29~AUD/kWh$ \citep{infinite_energy_synergy_2019} and increase at 4\% per annum, based on the average annual growth rate of Australian electricity prices over the previous 10~years \citep{abs_64010_2020}. The real options evaluation requires that at least one investment opportunity has a Discounted Payback Period of under 5~years for an investment to be made.

The value of the FiT plays a significant role in incentivising various configurations of household PV battery systems,\footnote{Generally higher FiTs accelerate the adoption of PV but delay the cost-effective tipping point of PV-battery systems. While lower FiTs initially reduce PV adoption, it also brings forward the tipping point for PV-battery systems that simultaneously drive further growth in additional PV capacity \citep{say_power_2019}.} thus three FiT scenarios are evaluated using time-invariant FiTs valued at 0\%, 25\% and 50\% of volumetric usage charges (i.e., it only applies to the quantity of excess PV generation exported to the network). This range is consistent with Australian retail FiTs in 2019 \citep{solar_choice_which_2019}. As is standard practice to maintain hosting capacity on the SWIS network \citep{synergy_am_2019}, a $5~kW_P$ FiT eligibility limit is used, such that PV systems above $5~kW_P$ lose all excess PV generation payments.\footnote{On other Australian networks, special approval is typically required to connect PV inverters greater than $5~kW$ to the grid \citep{solar_choice_solar_2019}.}  

\subsection{Input data for \textit{DIETER-WA}\label{subsec: input dieter-wa}}

The residual network demand profile used in \textit{DIETER-WA} is derived from SWIS network demand data provided by \cite{aemo_wem_2019} combined with the scaled net grid utilization of prosumage households as determined by \textit{Electroscape}. Historical time series of hourly wind power availability in the SWIS are provided by \cite{aemo_wem_2019-1}. To ensure utility-scale PV generation remains temporally consistent with household PV generation, the utility PV availability profile equals the average PV generation across each of the 261 households.

As for conventional utility-scale generation technologies, we include coal- and natural gas-fired plants, i.e.,~combined cycle gas turbines (CCGT) and open cycle gas turbines (OCGT), as well as bioenergy, onshore wind power, and utility PV. We further allow for investments in utility-scale batteries and hydrogen storage.\footnote{Under the parameterisation used here, we find that bioenergy and 'power-to-gas-to power' hydrogen storage are never part of the least-cost portfolio. We accordingly do not report on these technologies in the following.} Key techno-economic input parameters for these technologies are summarised in Table \ref{tab: inputs dieter-wa}.\footnote{The complete input data is available in the open-source spreadsheet provided with the model.} We ensure that both utility-scale and household PV and battery storage technologies utilise the same relative cost reduction curves mentioned in section \ref{subsec: input electroscape}. We also include a lower bound for wind power and utility PV investments corresponding to the capacity already in place \citep{aemo_aemo_2018}.

\begin{landscape}
\begin{table}[h]
\tiny
\begin{center}
\begin{tabular}{m{0.19\textwidth} l l l l l l l l l m{0.2\textwidth} }
\hline
\multirow{2}{0.19\textwidth}{\centering\textbf{Input parameter}}	&	\multirow{2}{4em}{\centering\textbf{Unit}}	&	\multirow{2}{4em}{\centering\textbf{Hard coal}}	&	\multirow{2}{4em}{\centering\textbf{CCGT}}	&	\multirow{2}{4em}{\centering\textbf{OCGT}}	&	\multirow{2}{4em}{\centering\textbf{Bioenergy}}	&	\multirow{2}{4em}{\centering\textbf{Wind power}}	&	\multirow{2}{4em}{\centering\textbf{PV}}	&	\multirow{2}{4em}{\centering\textbf{Li-ion storage}}	&	\multirow{2}{4em}{\centering\textbf{Hydrogen storage}}	&	\multirow{2}{4em}{\centering\textbf{Source}}	\\
    &   &   &   &   &   &   &   &   &   & \\
\hline
Overnight investment costs	&	AUD/MW	&	3,195,000	&	1,254,000	&	877,000	&	12,432,000	&	1,874,000	&	817,000	&	115848	&	2384615	&	\cite{graham_gencost_2018}, \cite{schmidt_future_2017}, \cite{pape_2014}	\\
Overnight investment costs (energy)	&	AUD/MWh	&	-	&	-	&	-	&	-	&	-	&	-	&	173773	&	308	&	\cite{schmidt_future_2017}, \cite{pape_2014}	\\
Annual fixed cost	&	AUD/MW	&	53200	&	10500	&	4200	&	131600	&	36000	&	14400	&	2027	&	16694	&	\cite{graham_gencost_2018}, own assumptions	\\
Variable OM costs	&	AUD/MWh	&	4.2	&	7.4	&	10.5	&	8.4	&	2.7	&	0	&	0.5	&	0.5	&	\cite{graham_gencost_2018}, own assumptions	\\
Thermal efficiency or roundtrip efficiency	&	\%	&	40	&	48	&	31	&	23	&	-	&	-	&	92	&	41.9	&	\cite{graham_gencost_2018}, \cite{pape_2014}	\\
Fuel costs	&	AUD/MWh$_{th}$	&	12.06	&	31.68	&	31.68	&	4.5	&	0	&	0	&		&		&	\cite{graham_gencost_2018}	\\
Technical Lifetime	&	years	&	25	&	25	&	25	&	25	&	25	&	25	&	15	&	22.5	&	\cite{graham_gencost_2018}, \cite{pape_2014}	\\
Lower bound for investment	&	MW	&	0	&	0	&	0	&	0	&	419	&	202	&	0	&	0	&	\cite{aemo_aemo_2018}	\\
\hline
\end{tabular}
\caption{Input parameters and data used in \textit{DIETER-WA}\label{tab: inputs dieter-wa}}
\end{center}
\end{table}
\end{landscape}

\section{Results\label{sec: results}}

\subsection{Changes in residual network demand from investments in PV battery prosumage\label{subsec: results residual demand}}

\begin{figure}[ht]
\centering{} \includegraphics[width=0.5\textwidth]{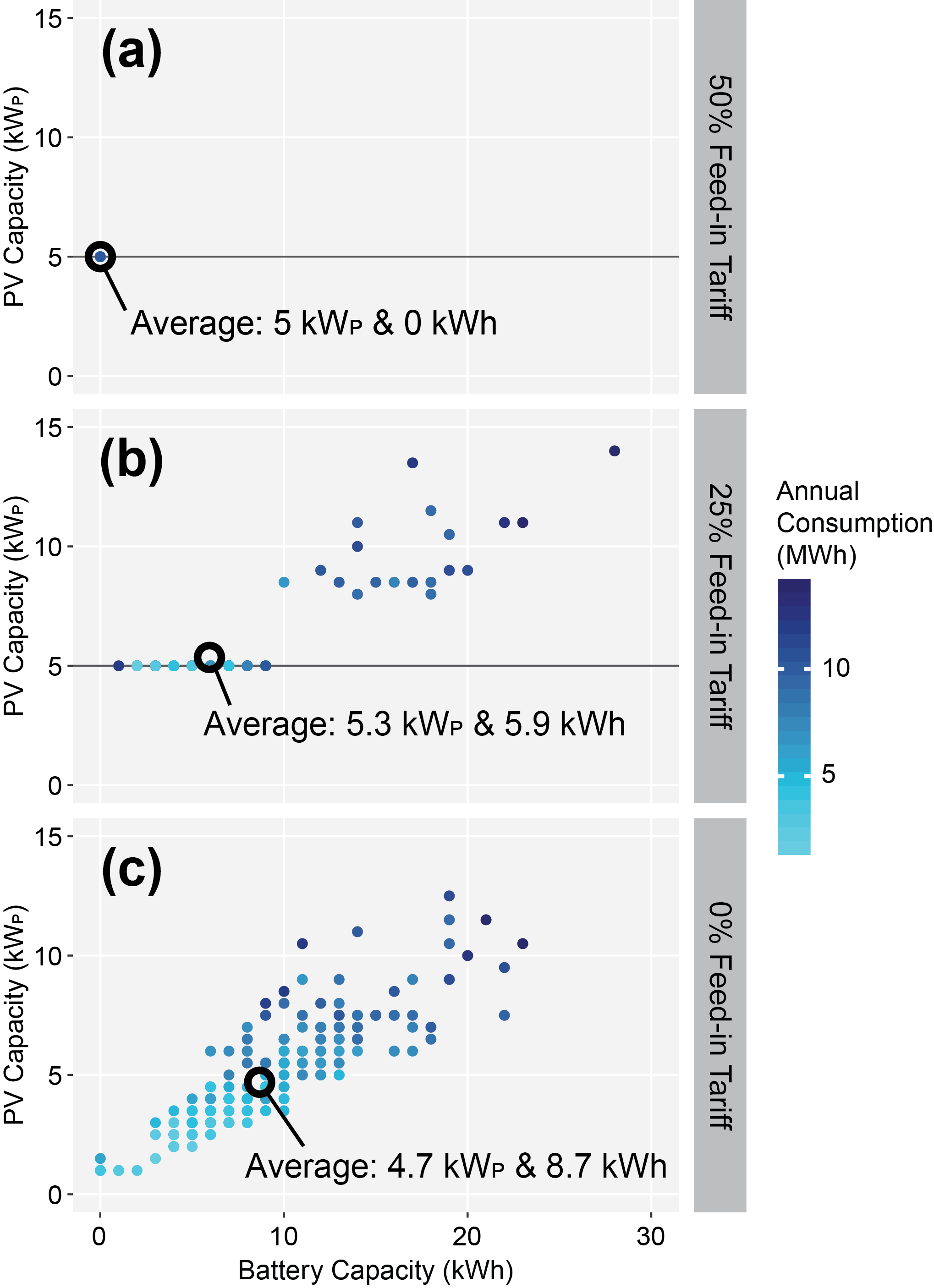}\\
\protect\caption{\label{fig: pv and battery capacity}Individual and average installed PV and battery capacities for each FiT scenario in the year 2030. (a) FiT valued at 50\% of volumetric usage charges, if $\leq 5~kW_P$. (b) FiT $\leq 5~kW_P$. (c) Without any FiT.}
\end{figure}

Investments by prosumage households in PV and battery capacity are heavily influenced by the value of the FiT (Figure \ref{fig: pv and battery capacity}). Higher FiTs provide greater returns for excess PV exports, encouraging larger PV systems (up to the $5~kW_P$ FiT eligibility limit), while lowering returns for self-consumption and discouraging the use of battery energy storage. As a result, the scenario with a FiT equalling 50\% of volumetric usage charges drives all 261~households by 2030 to invest in $5~kW_P$ PV systems with no battery storage. As the value of the FiT lowers, the value of self-consumption increases and the cost-effectiveness of battery storage is improved. In the scenario with a 25\% FiT, the increased value of self-consumption results in 7.3\% of households (with above average electricity consumption) foregoing their FiT revenue and installing PV systems above the $5~kW_P$ FiT eligibility limit. This raises the average PV capacity per household to $5.3~kW_P$ with $5.9~kWh$ of accompanying battery storage. In the scenario without a FiT (or 0\% FiT), the lack of financial incentive to export excess PV generation discourages excessively large PV systems while maximising the value of household self-consumption. This results in households investing in slightly smaller PV systems but with even larger battery capacities. The overall average PV capacity per household is $4.7~kW_P$ with $8.7~kWh$ of accompanying battery storage.\footnote{Qualitatively similar results have been found for other countries, e.g.~for Germany \citep{gunther_prosumage_2019}.}

Each of the FiT scenarios (0\%, 25\%, 50\% of volumetric usage charges) results in different average configurations of PV battery systems. To assist with readability, these three FiT scenarios will be respectively referred to as the ‘PVB+~FiT$_{0}$’, ‘PVB~FiT$_{25}$’ and ‘PV-only~FiT$_{50}$’ scenarios. 

The installed PV battery systems affect residual network demand by removing a household's load from the network (during self-consumption) and acting as a negative load (during excess PV exports). In the reference case without household PV battery investments, the annual residual network demand is $18.1~TWh$. Normalising the 261~households evaluated in \textit{Electroscape} to a single representative household and then scaling to $500,000$ households leads to the following reductions in annual residual network demand. In the ‘PV-only FiT$_{50}$’ scenario, with an average of $5~kW_P$ of PV and no batteries, the annual residual network demand is reduced to $15.1~TWh$ (or $-16.7$\%). In the ‘PVB FiT$_{25}$’ and ‘PVB+ FiT$_{0}$’ scenarios, the annual residual network demand is respectively reduced to $14.7~TWh$ ($-17.9$\%), and $15.2~TWh$ ($-15.6$\%). Since household PV generation is either self-consumed, exported, or time shifted (minus round-trip efficiency losses), these annual residual network demand reductions are predominantly driven by installed household PV capacity. 

\begin{figure}[ht]
\centering{} \includegraphics[width=1\textwidth]{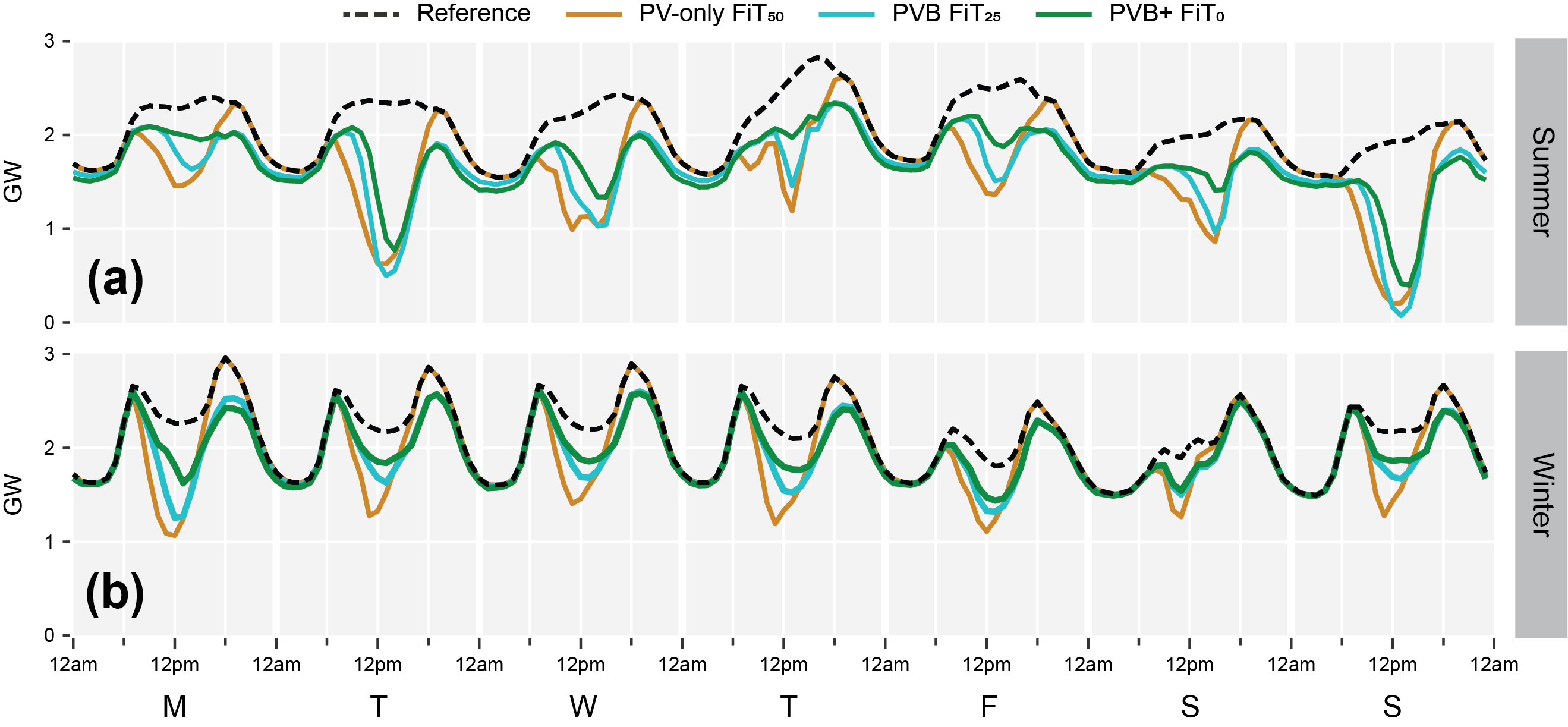}\\
\protect\caption{\label{fig: influence on residual network demand one week}Influence of the FiT scenarios on the SWIS residual network demand for 500,000 prosumage households across a week. (a) Week of the summer solstice (17 to 23 December). (b) Week of the winter solstice (18 to 24 June).}
\end{figure}

While annual residual network demand does not significantly differ between the FiT scenarios, their influence becomes much more evident at the diurnal scale (Figure \ref{fig: influence on residual network demand one week}). Generally, the minimum residual network demand each day begins to occur increasingly over midday due to the timing of excess PV generation. Potential reductions in the early evening peak depend upon the presence of a battery system. In the ‘PV-only FiT$_{50}$’ scenario, the peak residual network demand is delayed until sunset. In absolute terms, the diurnal peak demand can only be reduced slightly, and only during the summer months with long daylight hours. In the ‘PVB FiT$_{25}$’ and ‘PVB+ FiT$_{0}$’ scenarios, the household battery systems (that operate only to maximise self-consumption) are able to reduce peak residual network demand more strongly, and for a longer period of time. As it uses the PV generation stored during the day, larger battery systems (for a similar PV capacity) lead to a greater reduction of midday PV exports, thus reducing the down ramp of demand between the morning and midday, and the up ramp between midday and the early evening (comparing ‘PVB+~FiT$_{0}$’ and ‘PVB~FiT$_{25}$’ in Figure \ref{fig: influence on residual network demand one week} with ‘PV-only~FiT$_{50}$’).

\subsection{Impacts on optimal utility-scale generation and storage capacity\label{subsec: results capacity}}

\begin{figure}[ht]
\centering{} \includegraphics[width=1\textwidth]{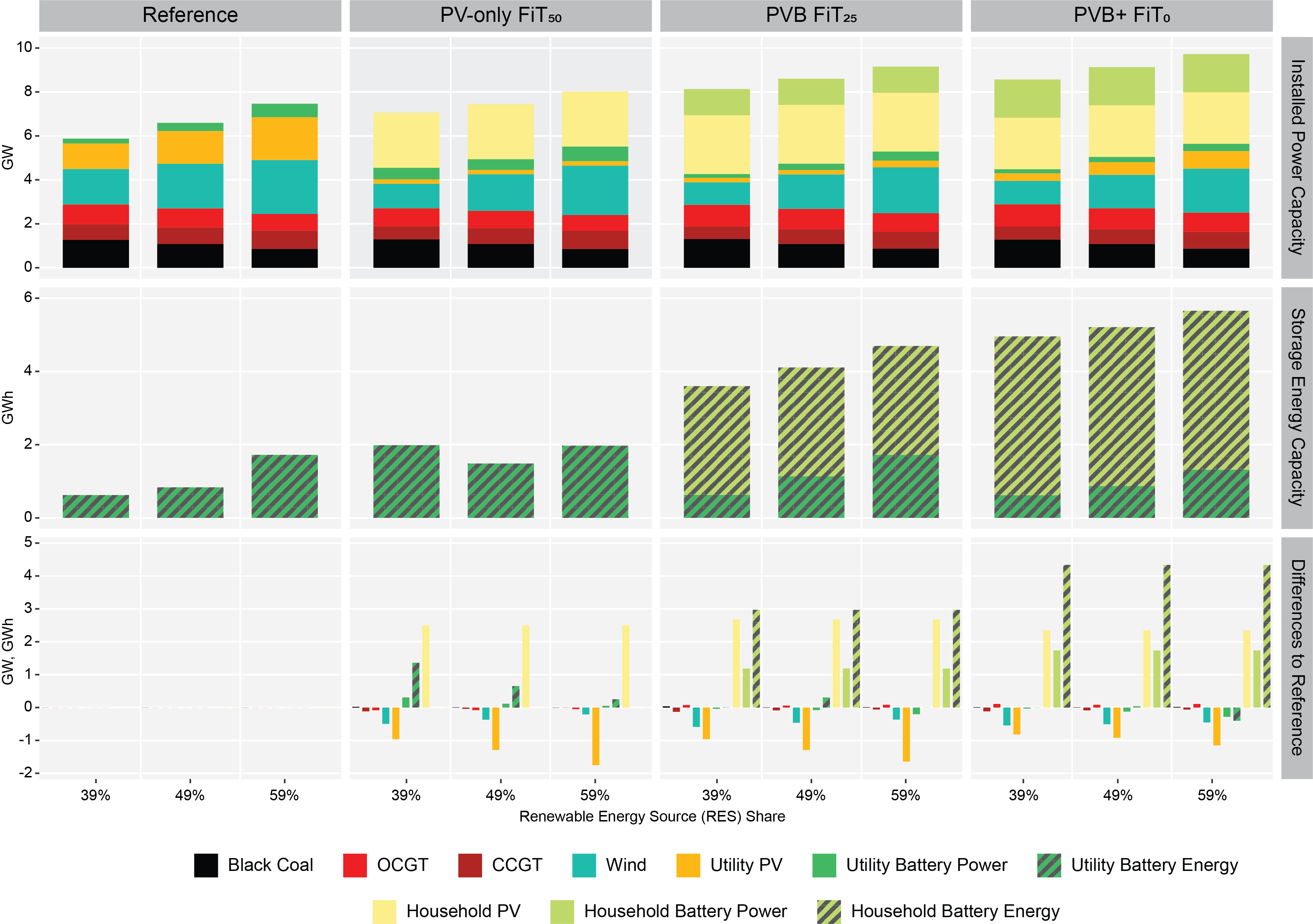}\\
\protect\caption{\label{fig: capacity changes}Installed power and storage energy capacity for varying FiT and RES shares (500,000 households) and the change in capacity with respect to the equivalent reference (i.e.,~without prosumage household investments) RES share scenario.}
\end{figure}

In the reference ‘39\% RES share’ scenario (i.e.,~without prosumage household investments), $1.16~GW_P$ of utility PV and $1.61~GW$ of wind power are optimal, along with relatively small utility battery storage of $0.21~GW$ and $0.62~GWh$ (Figure \ref{fig: capacity changes}, upper left panel). As the RES share rises to 59\%, utility PV and wind capacity increases to $1.95~GW_P$ and $2.45~GW$ respectively, while conventional generation capacity is reduced. The capacity of utility batteries raises disproportionately to $0.62~GW$ and $1.72~GWh$.

In the scenarios with prosumage, however, household PV capacity generally substitutes for utility-scale renewable energy generation capacity. The nature of this substitution depends upon the FiT values that incentivise different types of household PV-only or PV-battery investments that subsequently impact the timing of excess PV exports and the required contribution of utility-scale generation to the RES share.

In the ‘PV-only FiT$_{50}$’ scenario and across each RES share, utility PV experiences the largest reduction in capacity. Here, the cumulative PV capacity of PV-only prosumage households ($2.50~GW_P$) causes utility PV capacity to drop to the assumed lower bound of $202~MW_P$. In the 39\% RES share scenario, each $MW_p$ of household PV substitutes for $0.38~MW_P$ of utility PV and $0.20~MW$ of wind capacity, respectively. By generating at similar times, household PV capacity generally discourages additional utility PV capacity. As the RES share rises, relatively more utility PV and less wind power are substituted, as the respective utility-scale PV capacity in the reference is also larger. In the 59\% RES share scenario, each $MW_p$ household PV accordingly substitutes for $0.70~MW_P$ of utility PV and only $0.08~MW$ of wind power. The significant installed household PV capacity and absence of installed household battery systems in ‘PV-only~FiT$_{50}$’ also causes an increase of optimal utility battery power and energy storage capacity. This is because the increase in overall PV capacity and the corresponding decrease in wind power leads to larger diurnal variations between the midday and early evening residual network demand (compare Figure \ref{fig: influence on residual network demand one week}). This effect is particularly strong in the 39\% RES share scenario, which has the highest PV capacity share, with $0.12~MW$ and $0.55~MWh$ of additional utility battery capacity per $MW_p$ of household PV capacity. Conventional generation capacity hardly changes, except for a slight decrease in gas-fired generation capacity that corresponds to the increase in utility battery capacity.

In both the ‘PVB FiT$_{25}$’ and ‘PVB+ FiT$_{0}$’ scenarios, most effects are qualitatively similar. The substitution of utility-scale PV is slightly less pronounced because household battery systems partially balance the daily export of excess PV generation from household PV installations. Accordingly, this also slightly reduces the optimal amount of wind capacity. The only qualitative difference relates to utility batteries. Prosumage now slightly decreases the installed utility battery power capacity. Utility battery energy storage capacity, in contrast, remains constant or even increases. Overall, the substitution of utility batteries by household batteries is very incomplete, due to their operational focus on maximising self-consumption rather than wholesale energy arbitrage. Across RES shares, $1~MW$ of prosumage battery power capacity only substitutes for $0.02~MW$ to $0.17~MW$ of utility battery capacity; and $1~MWh$ of prosumage battery energy capacity substitutes for at most $0.09~MWh$ utility battery capacity (‘59\% RES share’ \& ‘PVB+~FiT$_{0}$’ scenario), but may also trigger an increase of up to $0.10~MWh$ (‘49\% RES share’ \& ‘PVB~FiT$_{25}$’ scenario). Conventional generation capacity again hardly changes, aside from a minor substitution between CCGT and OCGT capacity.

\subsection{Impacts on optimal yearly utility-scale generation and storage\label{subcec: results generation}}

\begin{figure}[ht]
\centering{} \includegraphics[width=1\textwidth]{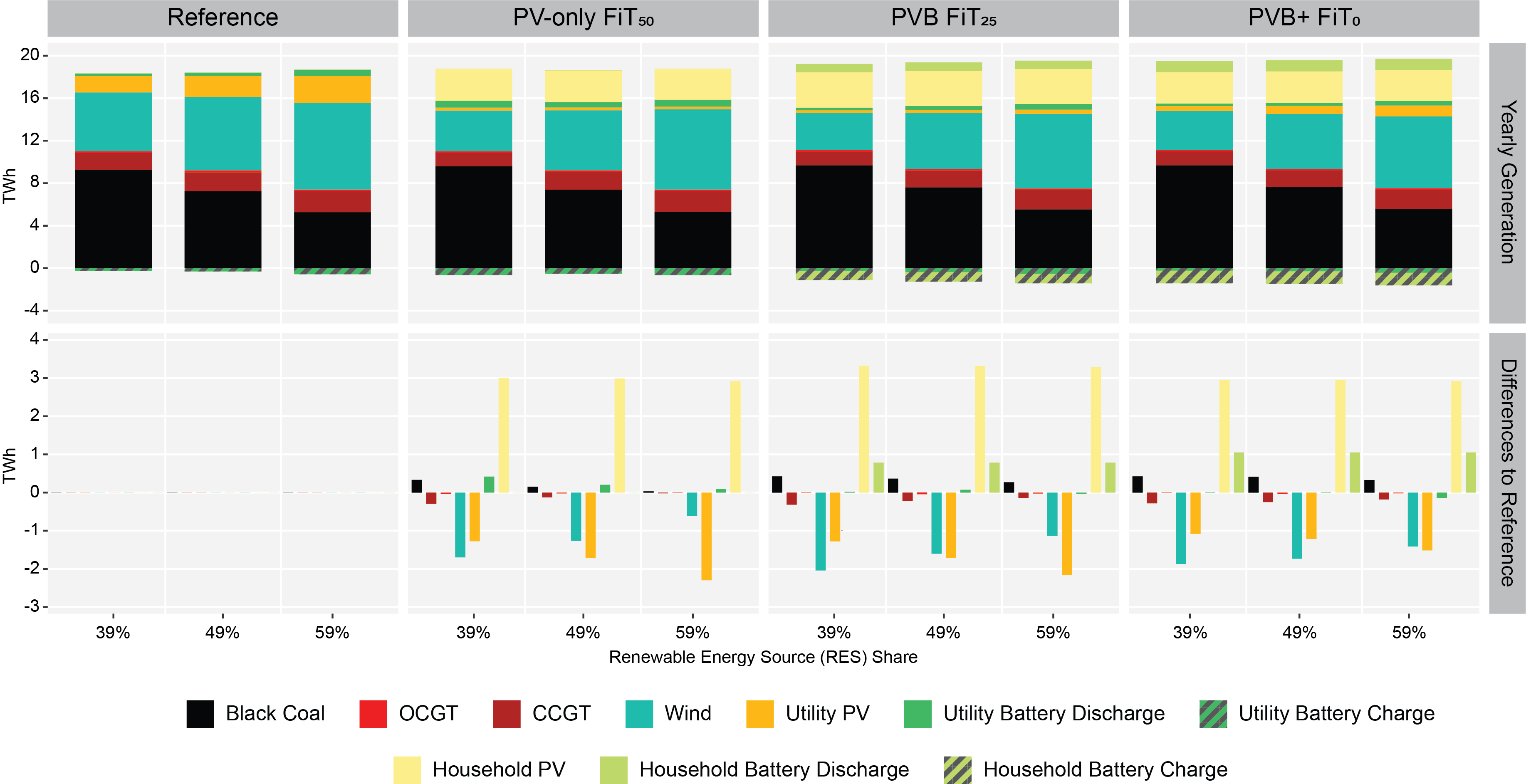}\\
\protect\caption{\label{fig: generation changes}Yearly generation for varying FiT and RES shares (500,000 households) and the change in generation with respect to the equivalent reference (i.e.,~without prosumage household investments) RES share scenario.}
\end{figure}

As the RES share rises in the reference scenario (upper left panel of Figure \ref{fig: generation changes}), wind power becomes an increasingly important resource in terms of yearly energy provided (30\% contribution at a 39\% RES share) and eventually begins to dominate the generation mix (45\% contribution at a 59\% RES share). The contribution of utility PV also slightly rises (9\% to 14\% between 39\% and 59\% RES shares). Coal generation has the greatest reduction (51\% to 29\%) while CCGT increases its share slightly, and OCGT generation remains generally unaffected.

In the scenarios with prosumage, wind generation generally experiences a larger overall reduction in terms of yearly generation when compared to the capacity effects described above, as wind power’s higher full load hours mean that capacity reductions have a larger energy impact. Raising the RES share again leads to a lower substitution of wind generation and a higher substitution of utility PV generation, slightly tempered with increasing deployment of household batteries (columns two, three and four of Figure \ref{fig: generation changes}).

\begin{figure}[ht]
\centering{} \includegraphics[width=0.6\textwidth]{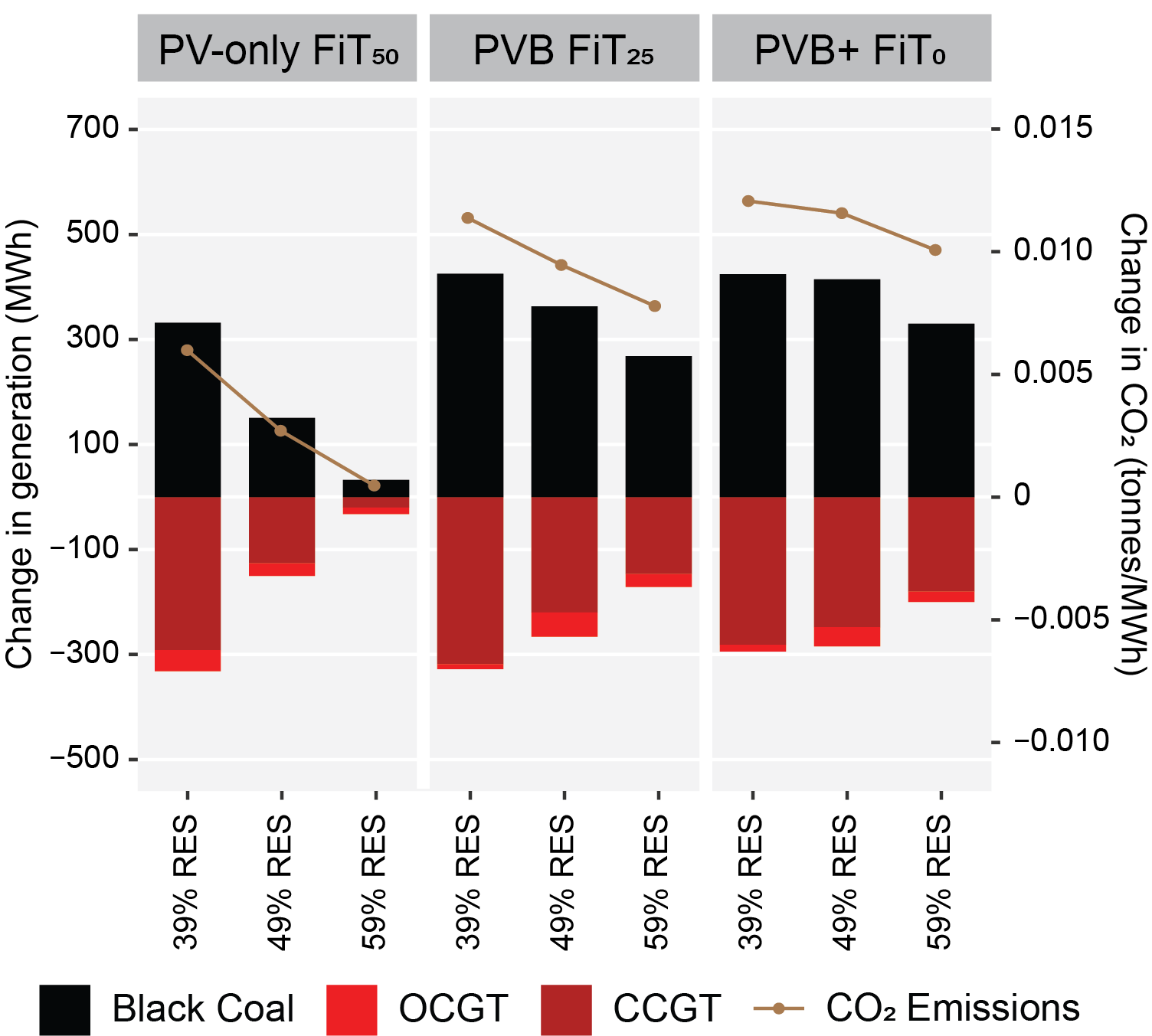}\\
\protect\caption{\label{fig: fossil generation changes}Change in power generation from coal and natural gas for varying FiT and RES shares and effects on CO$_2$ emissions per $kWh$ compared to reference scenario.}
\end{figure}

Overall power generation from coal increases slightly in the cases with prosumage (Figure \ref{fig: fossil generation changes}). This is most pronounced in the ‘PVB~FiT$_{25}$’ and ‘PVB+~FiT$_{50}$’ scenarios, where household batteries are also deployed. The increase in coal-fired power generation, combined with a corresponding decrease of natural gas-fired generation, also causes CO$_{2}$ emissions to slightly increase.\footnote{Additional model runs show that this finding disappears if the binding RES share constraint is relaxed.}

\begin{figure}[ht]
\centering{} \includegraphics[width=0.6\textwidth]{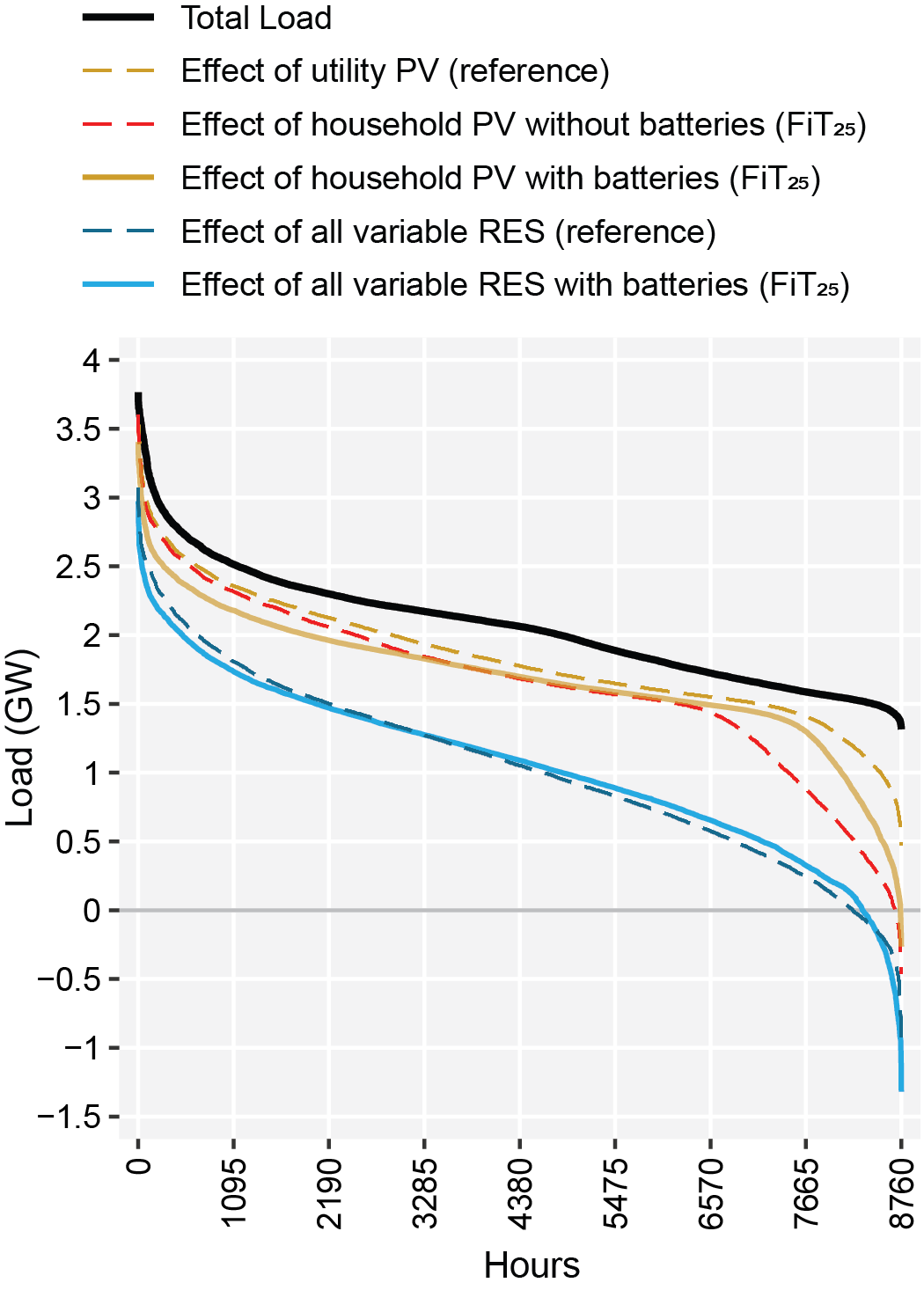}\\
\protect\caption{\label{fig: rldcs}Total load and residual load duration curves for the reference scenario and ‘PVB FiT$_{25}$’ for a RES share of 49\%.}
\end{figure}

Although the coal-enhancing effect of prosumage is small, an exploration of its drivers raises complementary insights. To do so, we look at Residual Load Duration Curves (RLDCs) of the reference scenario and the ‘PVB FiT$_{25}$’ scenario with a 49\% RES share (Figure \ref{fig: rldcs}).\footnote{For earlier and more detailed applications of residual load duration curves in the context of renewable energy integration and energy storage, see \cite{ueckerdt_2015} and \cite{zerrahn_sinn_2018}.} The blue curves show the residual load that remains to be served by utility-scale dispatchable generators and utility storage after the feed-in of all variable renewables. Here, the dashed blue line for ‘PVB FiT$_{25}$’ considers the net grid interaction of prosumage households, i.e.,~it takes not only household PV generation into account, but also the smoothing effect of behind-the-meter batteries. Comparing the two blue curves shows that PV-battery prosumage leads to an overall flatter residual load. While this is generally beneficial from a power sector perspective, it also allows coal-fired generators that have the lowest variable costs of non-renewable generators in our case study to slightly increase their production.

The changes in the residual load curve are driven (i) by an increasing overall solar PV capacity, and (ii) by the smoothing effect of household batteries. The dashed orange line shows the RLDC if only the utility-scale PV generation is taken into account in the reference scenario. The dashed red line then shows a counterfactual where this PV capacity grows to the size of prosumage household PV in the ‘PVB FiT$_{25}$’ scenario, but assuming that it would feed into the grid as utility-scale PV. That is, we counterfactually abstract from the smoothing effect of household batteries. The difference between these two curves thus shows the influence of having more PV in the power sector, triggered by prosumage. The solid orange curve then illustrates the smoothing effect of household batteries, i.e.~a decrease in residual load on the left-hand side, and an increase on the right-hand side.

Figure \ref{fig: rldcs} further shows a slightly increasing renewable surplus generation in the prosumage scenario on the very right-hand side. This is a consequence of the sub-optimal (i.e.,~oversized compared to the reference) overall PV capacity. It also explains why the optimal utility-scale storage energy capacity does not decrease further, or even increases, as this capacity is used for shifting surplus energy to other hours. Utility storage power, in turn, does not decrease further because it is still required for contributing to peak residual load coverage. Note that the RLDC hardly changes on the very left-hand side, i.e.,~the peak residual load stays high. We find that the utility-scale battery power is optimized such that it exactly covers the difference between the peak residual load and other dispatchable generators.

\subsection{Impacts on wholesale prices and system costs\label{subsec: results prices and costs}}

The use of household PV-battery systems not only benefits prosumage households (by significantly reducing the total amount of grid-imported energy) but also reshapes the overall residual network demand. This affects wholesale electricity prices and has implications for the cost of supplying electricity to different types of customers that have their own specific grid utilisation profiles. In cost minimisation modelling, the shadow price of a model’s energy balance is often interpreted as a wholesale electricity market price, which is for example used for market value analyses of renewable energy sources.\footnote{See \cite{brown_decreasing_2020} for a recent discussion on this.} We use this approach to calculate the weighted yearly average wholesale market prices for three customer types, using their respective grid utilisation profiles: (i) 500,000 prosumage households, (ii) 500,000 non-prosumage households,\footnote{As the Western Australian SWIS network has approximately 1~million residential customers at the end of 2018 \citep{aemo_2019_2019-1} and we have assumed 500,000 prosumager households previously.} and (iii) the remaining Commercial and Industrial (C\&I) demand. We assume both prosumage and non-prosumage households to have the same underlying grid demand as derived from \cite{ausgrid_solar_2018}.\footnote{In this setup, residential household demand equates to 31\% of total network demand and remains consistent with SWIS conditions in 2018 \citep{aemo_2019_2019-1}.} For clarity, we focus on the central ‘49\% RES share’ and ‘PVB FiT$_{25}$’ scenario.

\begin{figure}[ht!]
\centering{} \includegraphics[width=0.6\textwidth]{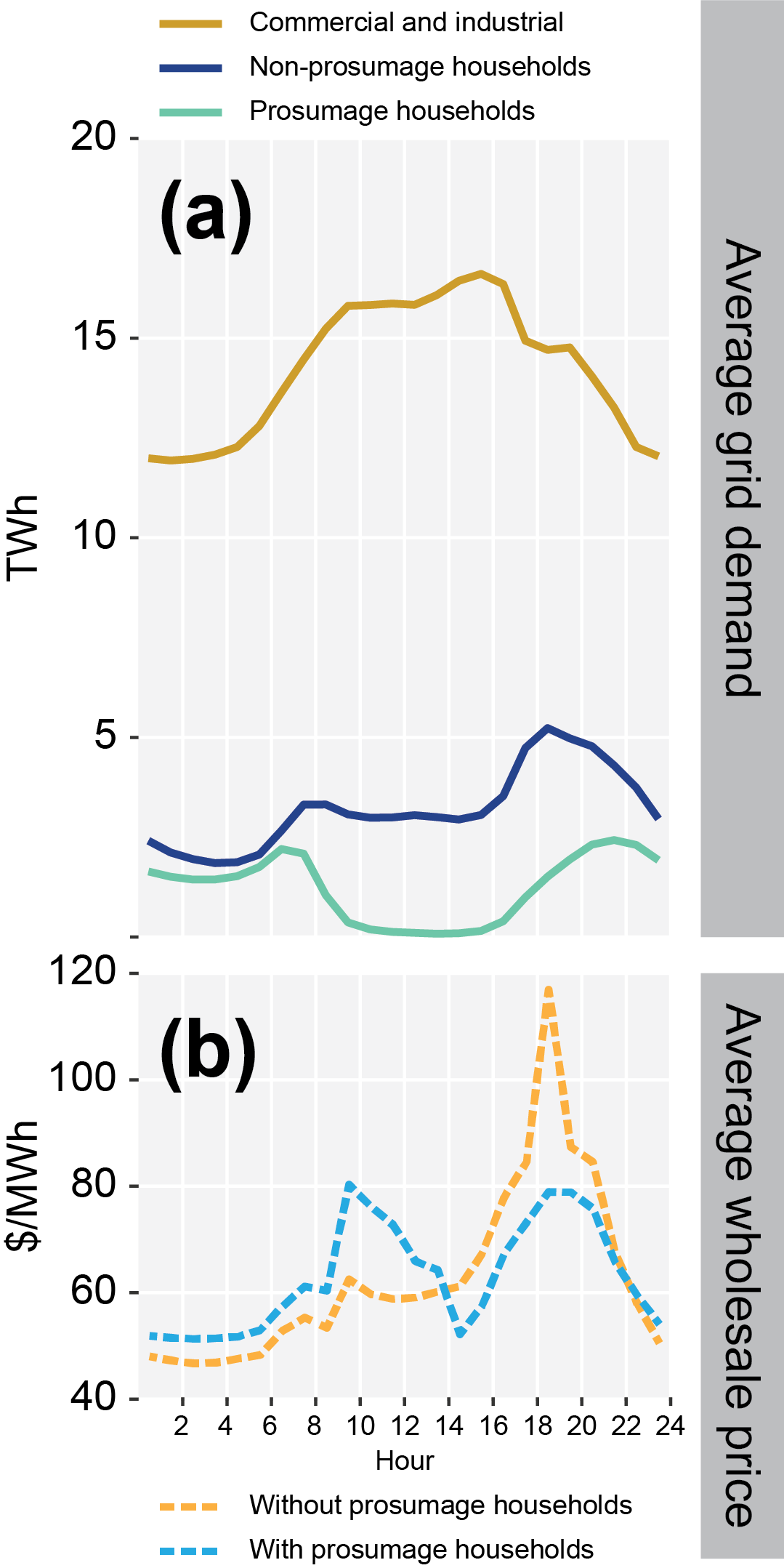}\\
\protect\caption{\label{fig: prices and grid utilisation}Effect of household PV-battery adoption in the ‘PVB FiT$_{25}$’ and ‘49\% RES share’ scenario on: (a) Average grid demand per hour for each customer segment. (b) Average wholesale electricity prices per hour.}
\end{figure}

By comparing the average hourly grid demand profiles (Figure \ref{fig: prices and grid utilisation}) between prosumage households, non-prosumage households and C\&I, it can be observed that residential households have a typical double peak profile (with a larger peak in the early evening, a much smaller peak in the morning and minimum demand during the night). With PV-battery systems, prosumage households on average are able to reduce the majority of their grid demand during the day with PV self-generation while continuing to reduce their grid demand past the early evening peak and partly into the night with their energy storage. C\&I demand has a relatively more constant profile with higher demand occurring over the day. In the absence of prosumage households, wholesale electricity prices are, on average, highest in the early evening as household peak demand overlaps with declining C\&I demand. With prosumage households,  the early evening price peak reduces significantly, while an additional price peak during the mid-morning emerges. Night-time prices are also increased slightly on average, due to the reduced contribution of zero marginal cost wind power in the overall portfolio. 

\begin{figure}[ht]
\centering{} \includegraphics[width=0.6\textwidth]{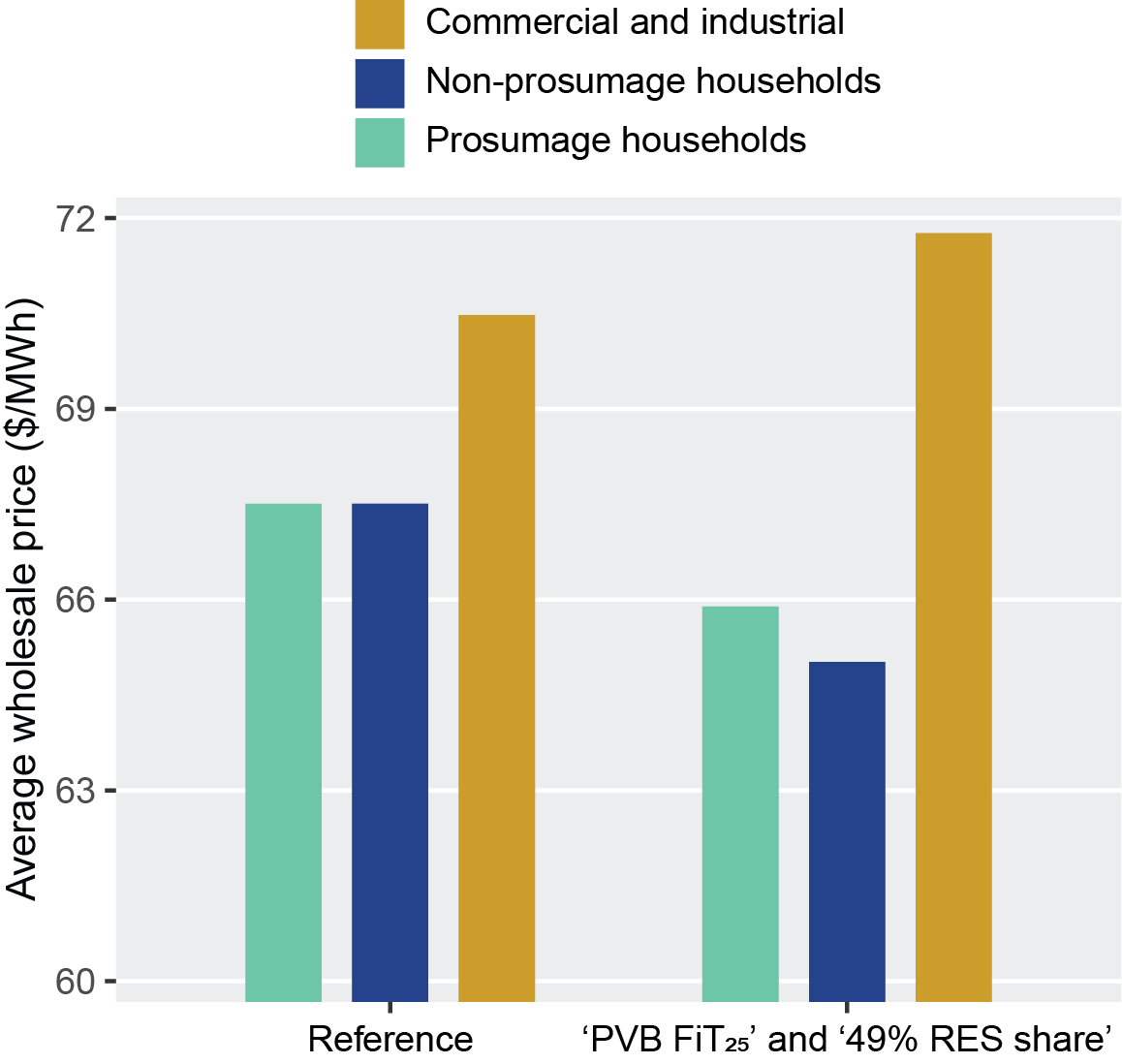}\\
\protect\caption{\label{fig: average prices}Comparison of average wholesale electricity prices for the three customer types with and without prosumage for the ‘49\% RES share’ and ‘PVB FiT$_{25}$’ scenario.}
\end{figure}

Average wholesale prices per $MWh$ of electricity decrease for prosumage households, but even more so for non-prosumage households, as they are able to benefit more from the large reduction in prices over their early evening peak (Figure \ref{fig: average prices}). Since prosumage households are generally discharging their battery systems during this time, they receive less advantage from this effect. But as prosumage households already benefit from significant reductions in their total net grid utilisation, they are able to obtain much greater overall wholesale market bill savings than non-prosumage households. Conversely, the wholesale prices for C\&I demand increase slightly, as they are relatively more exposed to the moderate rise of wholesale electricity prices in mid-morning and at night-time, and benefit less than households from the reduction of the evening price peak.

\begin{figure}[ht]
\centering{} \includegraphics[width=0.6\textwidth]{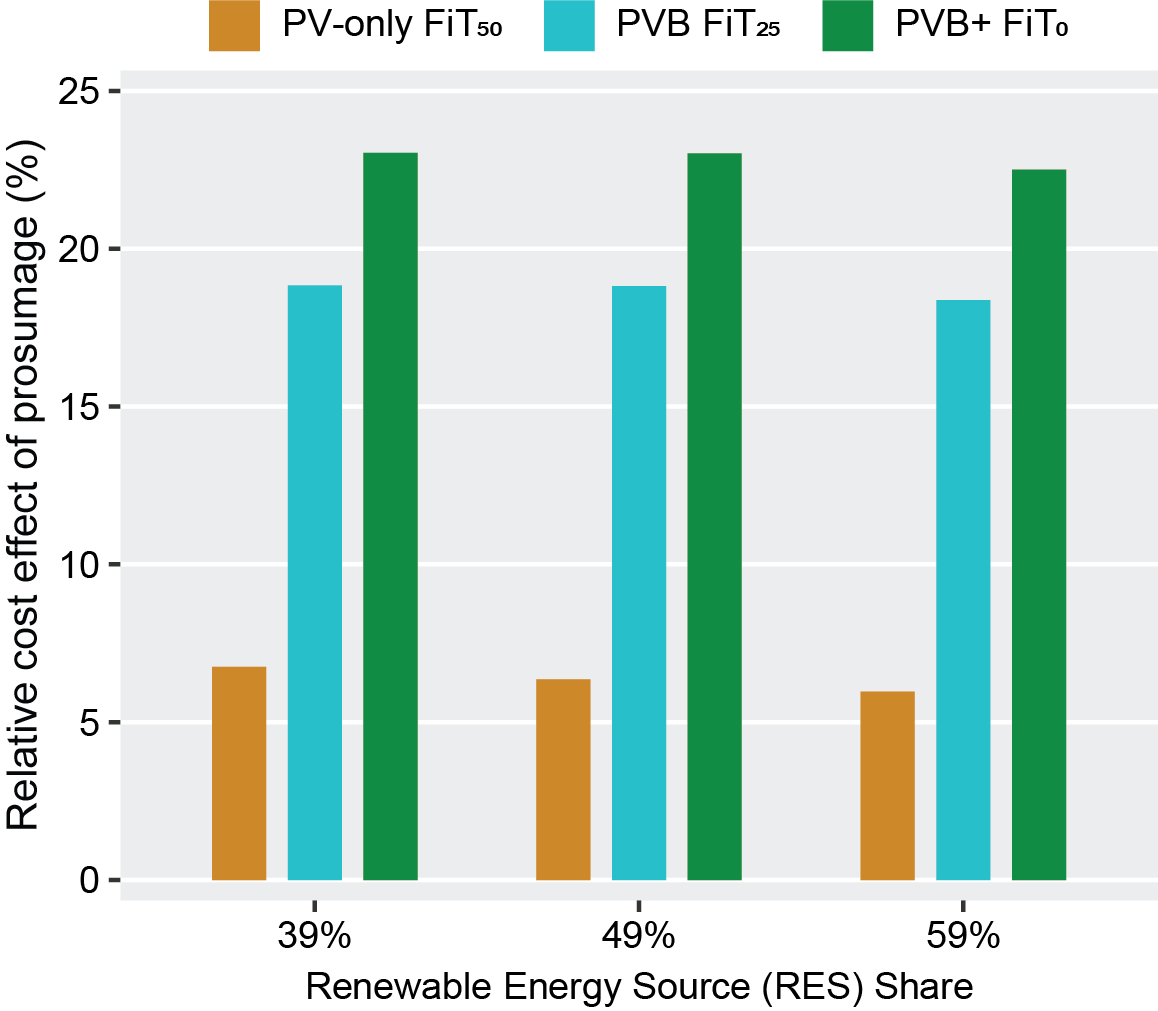}\\
\protect\caption{\label{fig: system costs}Relative effects of prosumage on overall system costs for each FiT and RES share scenario.}
\end{figure}

Overall system costs always increase with prosumage compared to the reference scenario (Figure \ref{fig: system costs}). This is mainly driven by additional battery deployment, and also by the fact that household investments in small-scale PV and battery capacity have higher specific investment costs compared to utility-scale investments. More generally speaking, prosumage households’ aim to reduce electricity bills by optimising self-consumption against the regulatory setting, which leads to a sub-optimal allocation of capital across the power sector. This is particularly visible in the ‘PVB FiT$_{25}$’ and ‘PVB+ FiT$_{0}$’ scenarios where installations of household batteries are high. Remember that these household batteries are not designed to assist with overall grid operation and thus hardly substitute utility-scale batteries. Additional information on system cost calculations are included in \ref{sub: appendix system costs}.

%

\section{Key assumptions and their qualitative impact on results\label{sec: key assumptions}}

In the following, we briefly discuss some key assumptions of the analysis and how these may qualitatively impact the results. 

First, our retail price assumptions are independent of the modelled power sector changes and wholesale market outcomes. This is a consequence of the research design that soft-links \textit{Electroscape} with \textit{DIETER-WA}. In \textit{Electroscape}, households’ PV battery investments are driven by an assumed constant increase in retail electricity prices that is independent of the renewable energy share and the number of prosumage households. This assumption mirrors a long-running empirical trend in Australia \citep{abs_64010_2020}. If retail price increases were lower than the 4\% per year assumed here, the uptake of household PV-battery systems would be slower and the accompanying system effects would be smaller (and vice versa). Increasing RES shares generally implies a greater penetration of zero-marginal cost generators, such that the wholesale cost of electricity may also decrease. Yet these savings may not be reflected in retail electricity prices. Wholesale electricity costs contribute to less than half of retail charges \citep{aemc_2019_2019}, and an increasing penetration of variable renewable energy often requires significant transmission and distribution network upgrades, that (though not explicitly modelled here) are usually recovered via increases in retail prices. Likewise, the financing of renewable energy remuneration schemes may also lead to further increases in retail prices. Accordingly, it appears justified to assume that the recent trend of increasing retail prices continues even if RES shares increase. Future research may aim to relax this assumption by further integrating both model approaches.

Second, we assume that FiTs are only eligible for PV system installations under $5~kW_P$. This mirrors the conditions in the SWIS network \citep{synergy_am_2019}, with similar arrangements occurring in other Australian networks \citep{solar_choice_solar_2019}. The $5~kW_P$ threshold is likely to become even more prevalent amid concerns with reverse power flows \citep{csiro_and_energy_networks_australia_electricity_2017} and revised inverter standards \citep{stapleton_overview_2017}. Higher thresholds would result in higher household PV capacities being installed, and accordingly larger power sector effects, particularly in the ‘PV-only FiT$_{50}$’ scenario.\footnote{Compare, for example, the German setting modelled by \cite{gunther_prosumage_2019}, where household PV installations are always at a $10~kW_P$ threshold whenever the FiT is sufficiently high.}

Third, we adopt a long-run equilibrium 'greenfield' perspective in \textit{DIETER-WA}.\footnote{It is not a pure greenfield assumption, as we include a small lower bound for utility wind and solar power, and the PV lower bound is binding in some cases ($202~MW$).} Accordingly, the optimal solution accounts for the full costs of all generators, both fixed and variable. In reality, existing conventional plants may not have to recover their full costs, but only the costs of going forward with the existing capacity. We may therefore underestimate the capacity of coal-fired power plants, which come with relatively high fixed costs, and conversely overestimated the capacity of OCGT plants. At the same time, an opposite distortion may be present, as we implicitly assume that the overall setting remains stable over time in our 2030 parameterisation. In the real world, investors may expect an ongoing transition toward higher shares of renewable energy sources or tighter carbon constraints beyond 2030, such that investments in coal- or gas-fired plants could be lower.

We further assume that certain shares of renewable energy sources are exactly met by including respective binding conditions in the model. This allows for meaningful comparisons of different scenarios and is relevant from an energy policy perspective, as relative renewable energy targets are common in many countries. Yet we force the model to deviate from an endogenous, cost-optimal renewable share. To explore this, we carry out additional model runs with an endogenous share of renewable energy sources. Given our parameterisation, we find an optimal\footnote{Note that this is not the optimal share of renewable energy sources from a social welfare perspective, i.e.,~if all external costs were internalised.} share of renewables just below 59\%, which slightly decreases with higher FiT values. Qualitatively, most results do not change compared to our setting with exogenously fixed renewable shares. Yet the coal-enhancing effect of prosumage described earlier disappears, as prosumage always increases the renewable penetration compared to the reference scenario, and accordingly substitutes more electricity from coal compared to the scenarios with fixed renewable shares.

Next, we abstract from technical limitations or additional costs related to ramping up and down power generation from conventional plants between one hour and the next in the \textit{DIETER-WA} model used here. Thus, we may have not fully captured the potential system benefits of the ‘PVB+ FiT$_{0}$’ scenario, which generally leads to a smoother residual load duration curve and lower ramps for thermal generators.

We further abstract from including a CO$_{2}$ price. While this is a meaningful policy assumption for the Western Australian case modelled here, it somewhat limits the interpretability of results for other jurisdictions where CO$_{2}$ pricing is present. In case a sufficiently high CO$_{2}$ price was introduced, coal-fired power plants would be substituted by natural gas. Accordingly, the minor coal-enhancing effect of prosumage would also disappear. Yet overall results are unlikely to change as the share of renewable energy sources is by assumption fixed.

Finally, our research design ignores the possibility that residential batteries could be used for further market or grid services, rather than only increasing the self-consumption of households’ PV generation. While this adequately reflects the current setting in Western Australia and many other markets, household batteries may increasingly become available for additional uses in the future, enabled by aggregators and new technologies. If residential batteries become available for such applications, they may substitute utility-scale storage to a greater extent and thus mitigate the overall system cost increase from prosumage. Exploring the potentials and preconditions for this in more detail appears to be a promising avenue for future research.

\section{Discussion and conclusions\label{sec: conclusions}}

Using two open-source models, we first determine optimal investments into residential PV and battery capacities from a financial household perspective and then analyse their wider power sector effects. Using different FiT values and RES shares, we illustrate how prosumage changes the residual network demand and overall utility-scale generation and storage capacity investments and dispatch. We do so for the Western Australian SWIS island network, which serves as an example of what many other countries may experience in the future. Accordingly, the following general outcomes, which are evident across the range of scenarios and results, should also be of interest to many other geographical settings.

First, residential PV generally displaces utility-scale PV capacity. This effect is less pronounced if more residential batteries are deployed, and more pronounced for higher RES shares. Accordingly, future investments in utility PV capacity will have to consider the growth of prosumage as it directly competes against their market dispatch. Therefore, the use of utility PV capacity in the future may require additional financial certainty by engaging in hedging agreements, such as contracts-for-difference, rather than relying solely upon market dispatch revenues from the wholesale energy market.

Second, the optimal wind capacity, in contrast, is generally less affected by prosumage. As self-generation by prosumage households contributes to the RES share, it naturally displaces the remaining share of renewable energy required from utility-scale generators. Across each of the FiT and RES share scenarios however, reductions in wind capacity are less pronounced than with utility PV capacity. Furthermore, raising the RES share drives additional wind capacity over utility PV capacity. From a central planner perspective, investments in wind capacity may be more resilient to different degrees of prosumage adoption.

Third, even if a substantial residential battery capacity is deployed, utility battery power capacity is only displaced slightly while utility battery energy capacity may even increase. There is therefore a very imperfect substitution of utility storage by residential batteries. This is, amongst other factors, driven by the fact that we have assumed prosumage batteries to only be operated with the objective of minimising the households’ energy costs. This is also a major source of increasing overall system costs.

Fourth, prosumage causes average wholesale prices to slightly decrease for both prosumage and non-prosumage households and slightly increase for other consumers. These distributive implications of prosumage may be of interest from a policy perspective; yet the overall effects are also dependent on the design (and potential design changes) of retail tariffs and the pass-through by retailers to consumers.

Overall, we conclude that prosumage can have substantial impacts on the overall power sector, which has to be considered by system planners, investors and regulators alike. Such power sector repercussions should also be taken into account when designing FiTs, retail tariffs and self-consumption regulation for residential electricity customers that influence subsequent household PV-battery investment behaviours. Likewise, an increasing uptake of prosumage presents a potential opportunity to both better incorporate and utilise these behind-the-meter PV battery systems as a source of additional power sector flexibility.

Future work may explore in more detail the distributive impacts of prosumage and potential grid tariff reform options to mitigate these. Moreover, a further integration of the two models appears desirable, which would also allow the incorporation of retail price feedbacks from increased prosumage as well as investigating the effects of making additional use of prosumage batteries for grid storage purposes.

\section*{Acknowledgements}
We thank researchers from the Energy Transition Hub and the participants of the Climate \& Energy College seminar at the University of Melbourne for fruitful discussions. Kelvin Say acknowledges the resources provided by The Pawsey Supercomputing Centre with funding from the Australian Government and the Government of Western Australia. Wolf-Peter Schill acknowledges funding by the Federal Ministry of Education and Research via the START project (FKZ 03EK3046E) and the invitation for a research stay at the University of Melbourne, where parts of this work have been carried out.

\appendix

\section{Additional results}


\subsection{System costs\label{sub: appendix system costs}}

Although not in the focus of this analysis, we also compare overall system costs. As for all non-prosumage parts of the power sector, this is straightforward, as respective fixed and variable costs are direct outcomes of the \textit{DIETER-WA} model. On the prosumage side, we re-calculate the costs of investments in household PV and battery capacity determined by \textit{Electroscape} in a way that they are comparable to system cost calculations in \textit{DIETER-WA}. We do this by summing up the annuities of respective investments in every year between 2019 and 2030, using a discount rate of 4\% (the same as for utility-scale assets in \textit{DIETER-WA}) and lifetimes of 10 years for batteries and 25 years for PV installations. In doing so, we consider the higher specific investment costs of household PV and battery installations compared to their utility-scale counterparts.

\begin{figure}[ht]
\centering{} \includegraphics[width=1\textwidth]{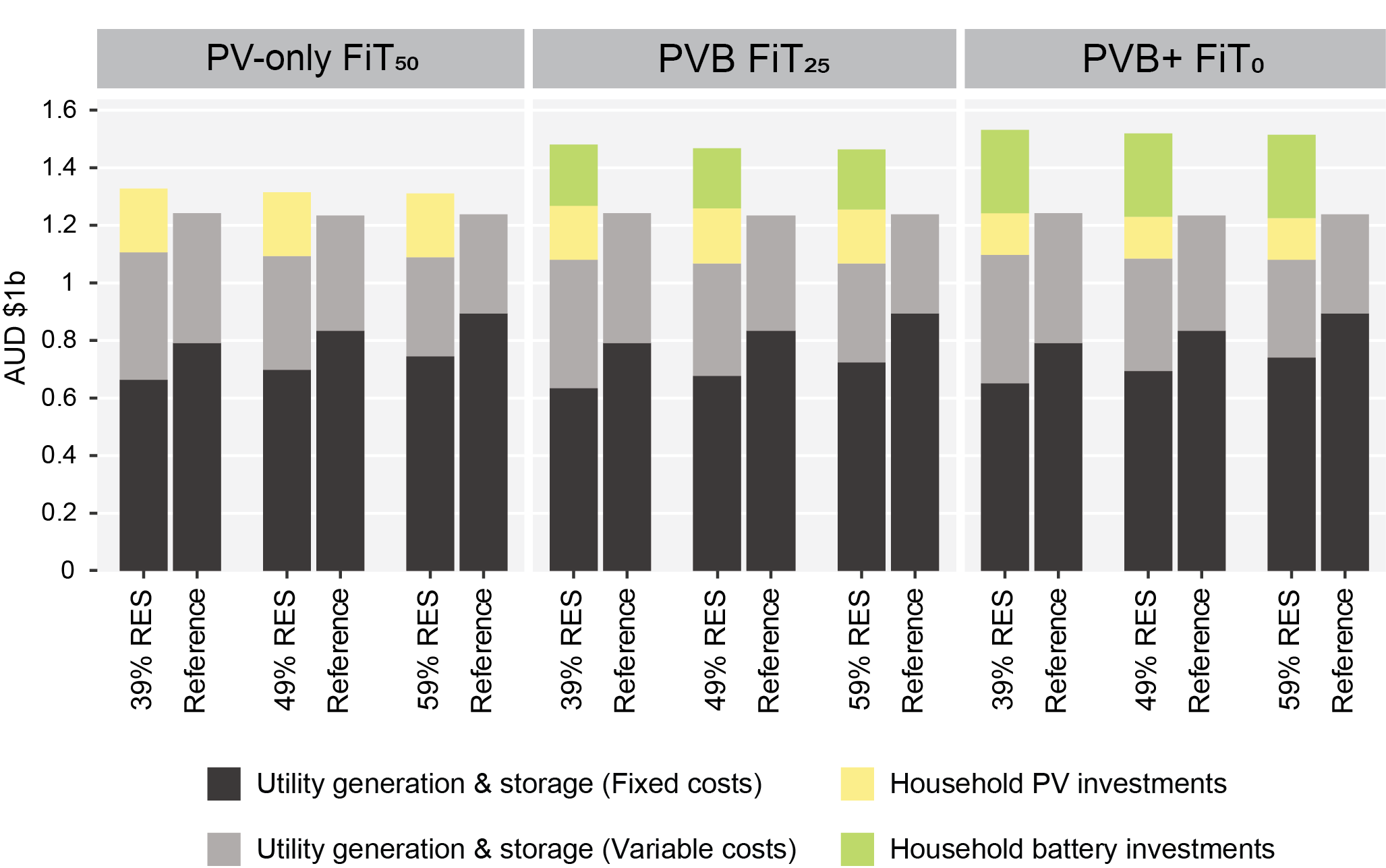}\\
\protect\caption{\label{fig: system costs appendix}Comparison of total annual system costs.}
\end{figure}

We find that overall system costs in the scenarios with prosumage are always higher than in the reference scenario. Generally speaking, this is because the inclusion of household PV battery systems forces the model to deviate from the least-cost generation and storage portfolio achieved in the reference scenario. In particular, prosumage batteries substitute utility battery storage only to a minor extent (compare Section \ref{subsec: results capacity}), so overall battery-related investments increase substantially.\footnote{Similar findings have been made for prosumage scenarios for Germany \citep{schill_prosumage_2017}.} Accordingly, system costs increase most in the FiT$_0$ scenario, where we find the highest prosumage battery investments (Figure \ref{fig: system costs appendix}). Depending on the renewable share, yearly system costs increase by around 23\% in the FiT$_0$ scenario compared to the reference, but only by 6-7\% in the FiT$_{50}$ scenario without batteries. Another factor that contributes to increasing overall system costs relates to higher specific investment costs of households’ PV and battery installations compared to their utility-scale counterparts.

\section{Research Data}

Source code and input data are publicly accessible from Zenodo repositories \url{http://doi.org/10.5281/zenodo.3693308} for \textit{Electroscape} and \url{http://doi.org/10.5281/zenodo.3693287} for \textit{DIETER-WA}.



\bibliography{mybibfile}

\newpage

\section{Sensitivity analyses}

\subsection{Solar PV system costs ($\pm~20\%$)}

\begin{figure}[h!]
\centering{} \includegraphics[width=0.82\textwidth]{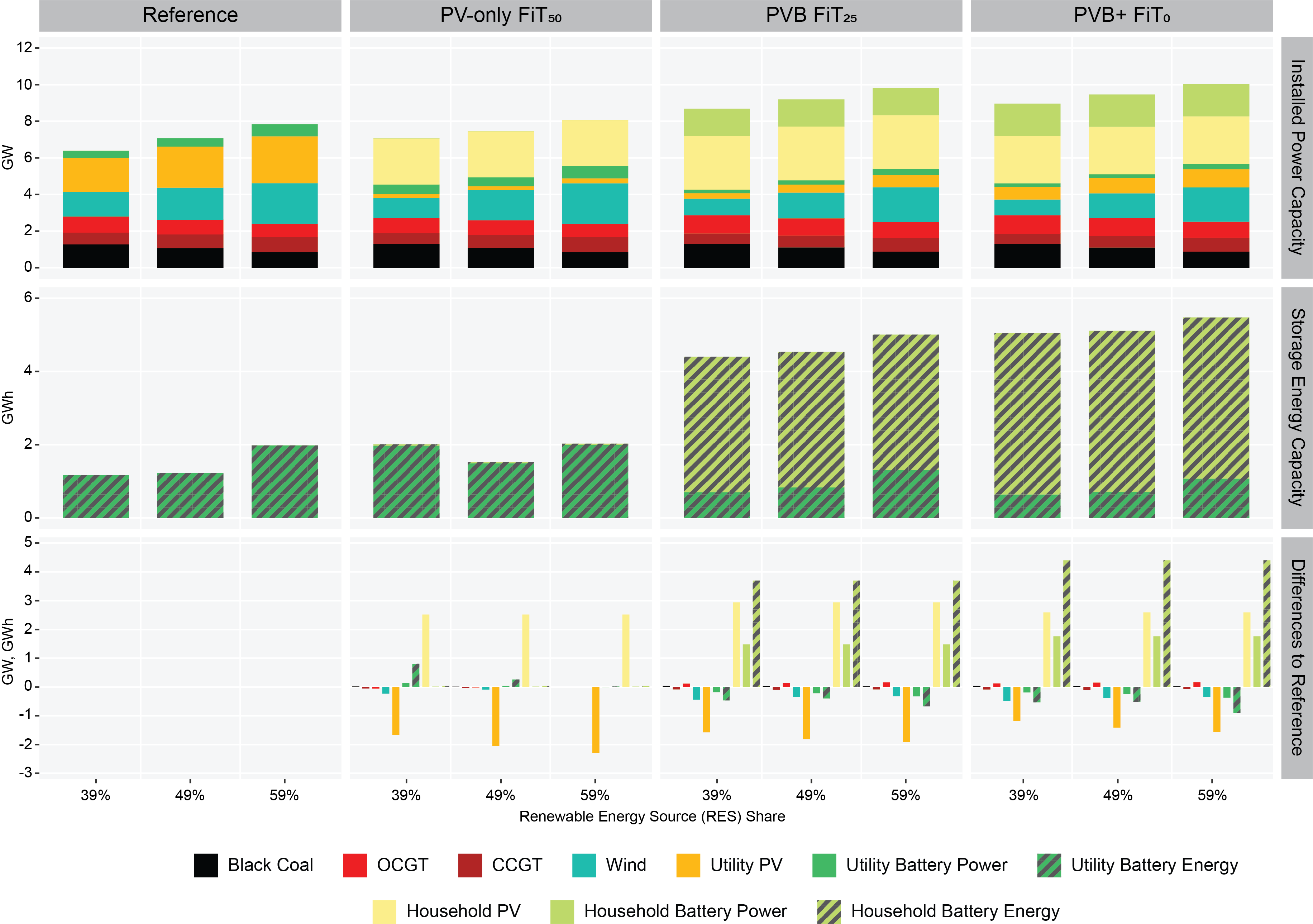}\\
\protect\caption{\label{fig: capacity changes pv20m} Installed power and storage energy capacity for varying FiT and RES shares (500,000 households) and the change in capacity with respect to the equivalent reference scenario, sensitivity with -20\% PV system costs.}
\end{figure}

\begin{figure}[h!]
\centering{} \includegraphics[width=0.82\textwidth]{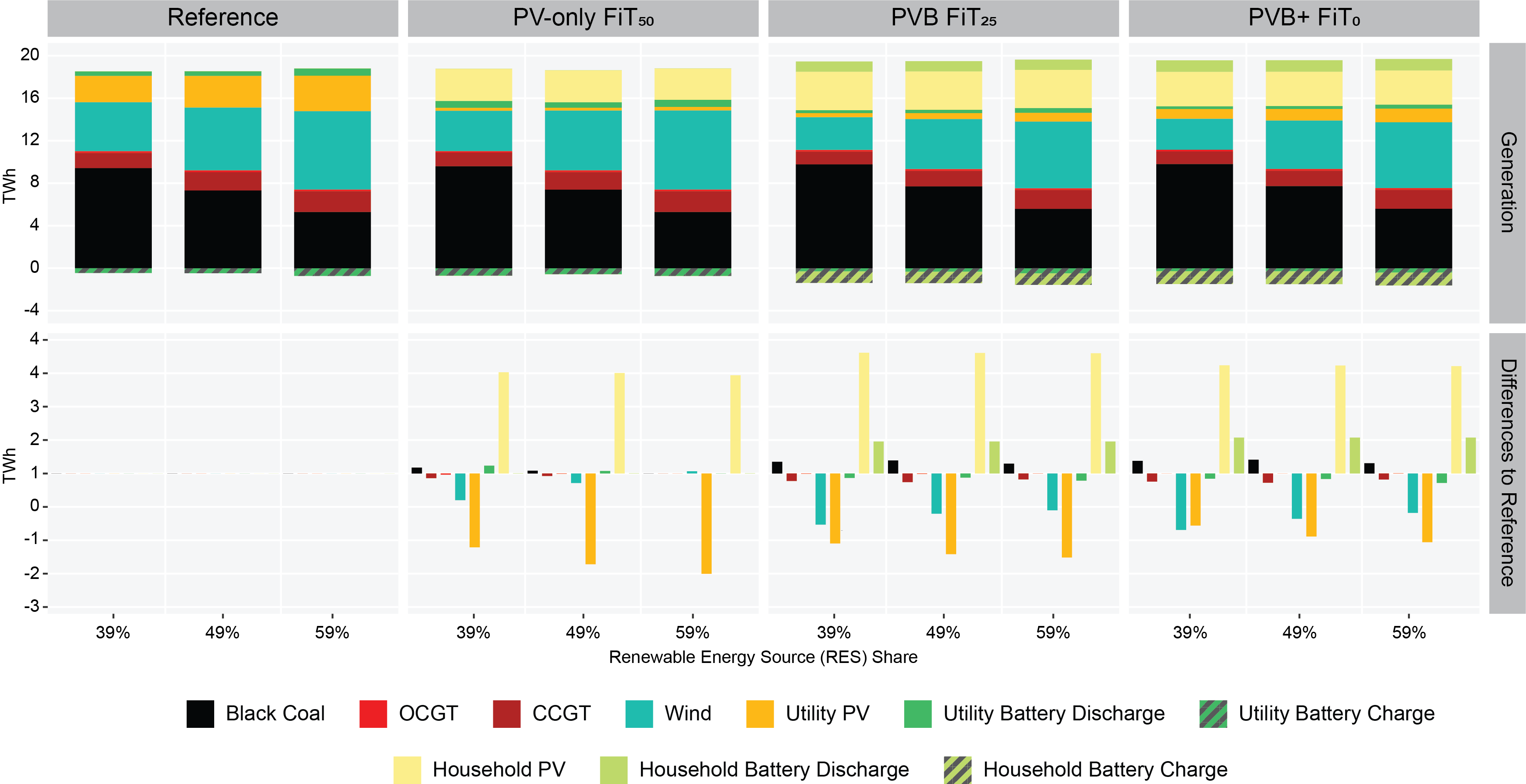}\\
\protect\caption{\label{fig: generation changes pv20m} Yearly generation for varying FiT and RES shares (500,000 households) and the change in generation with respect to the equivalent reference scenario, sensitivity with -20\% PV system costs.}
\end{figure}

A 20\% reduction in PV system costs generally raises utility and household PV capacity and reduces wind capacity compared to baseline assumptions (Figure \ref{fig: capacity changes pv20m}). Household PV continues to displace utility PV, but to a slightly lesser extent. Wind continues to remain the dominant source of renewable energy generation, but at a slightly lower level (Figure \ref{fig: generation changes pv20m}). Qualitatively, the impacts of household PV battery prosumage on utility-scale capacities and generation remain consistent.

\pagebreak

\begin{figure}[h!]
\centering{} \includegraphics[width=0.82\textwidth]{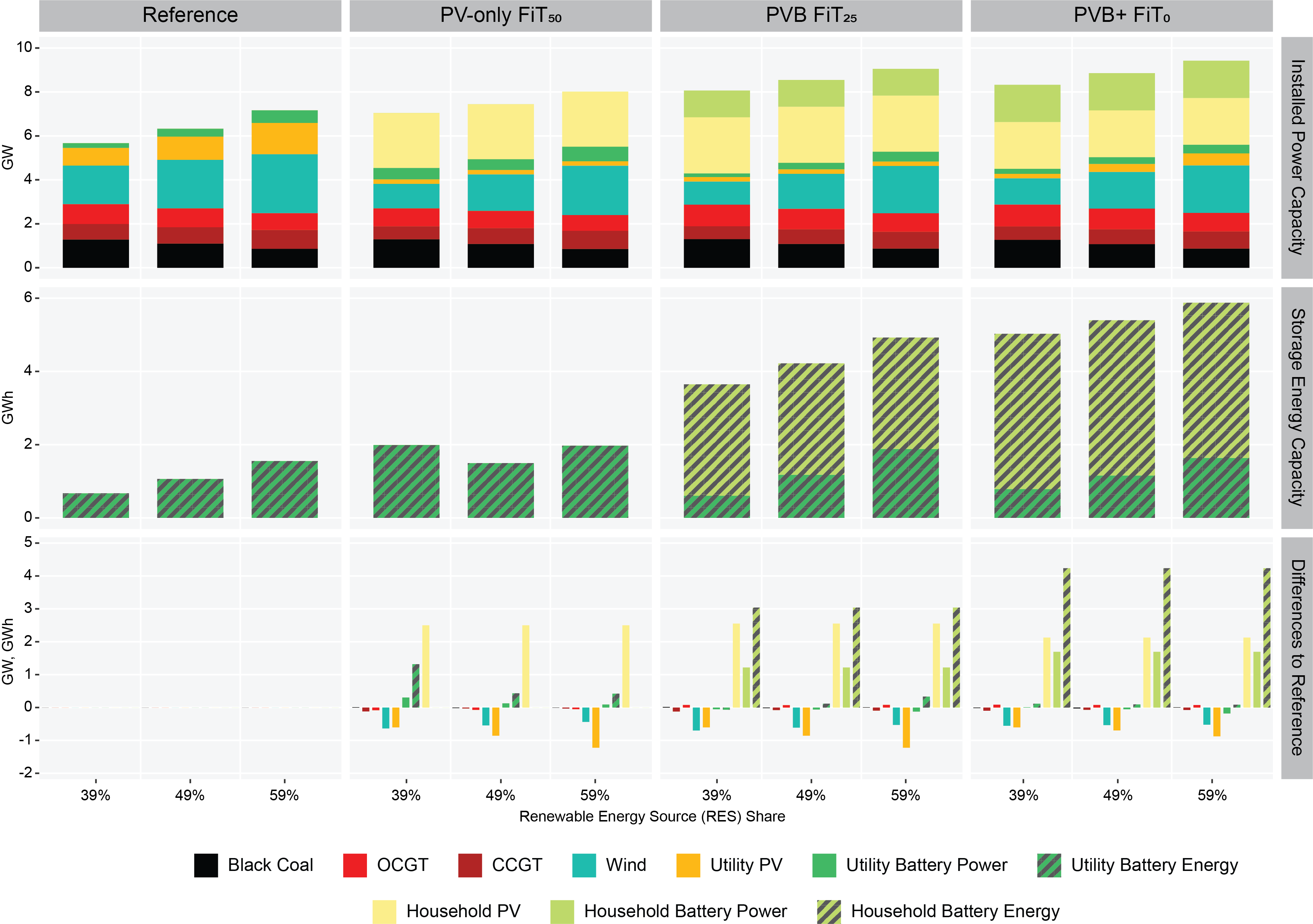}\\
\protect\caption{\label{fig: capacity changes pv20p} Installed power and storage energy capacity for varying FiT and RES shares (500,000 households) and the change in capacity with respect to the equivalent reference scenario, sensitivity with +20\% PV system costs.}
\end{figure}

\begin{figure}[h!]
\centering{} \includegraphics[width=0.82\textwidth]{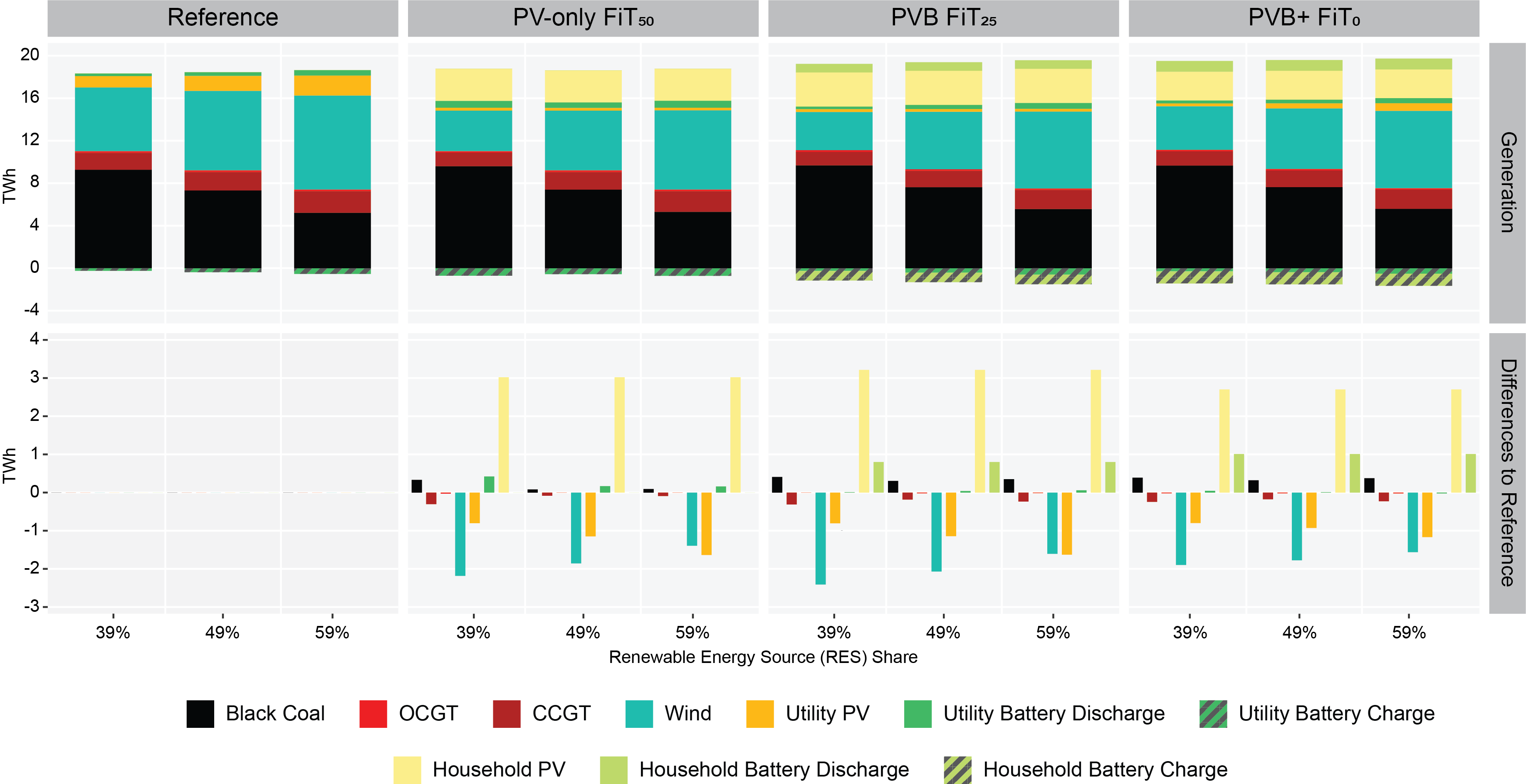}\\
\protect\caption{\label{fig: generation changes pv20p} Yearly generation for varying FiT and RES shares (500,000 households) and the change in generation with respect to the equivalent reference scenario, sensitivity with +20\% PV system costs.}
\end{figure}

A 20\% increase in PV system costs generally reduces utility and household PV capacity and increases wind capacity compared to baseline assumptions (Figure \ref{fig: capacity changes pv20p}). With household battery capacity remaining stable, the higher proportion of wind generation (Figure \ref{fig: generation changes pv20p}) drives a general increase in utility battery capacity. Qualitatively, the power sector implications from household PV battery prosumage remain consistent.

\pagebreak

\subsection{Battery storage system costs ($\pm~20\%$)}

\begin{figure}[h!]
\centering{} \includegraphics[width=0.82\textwidth]{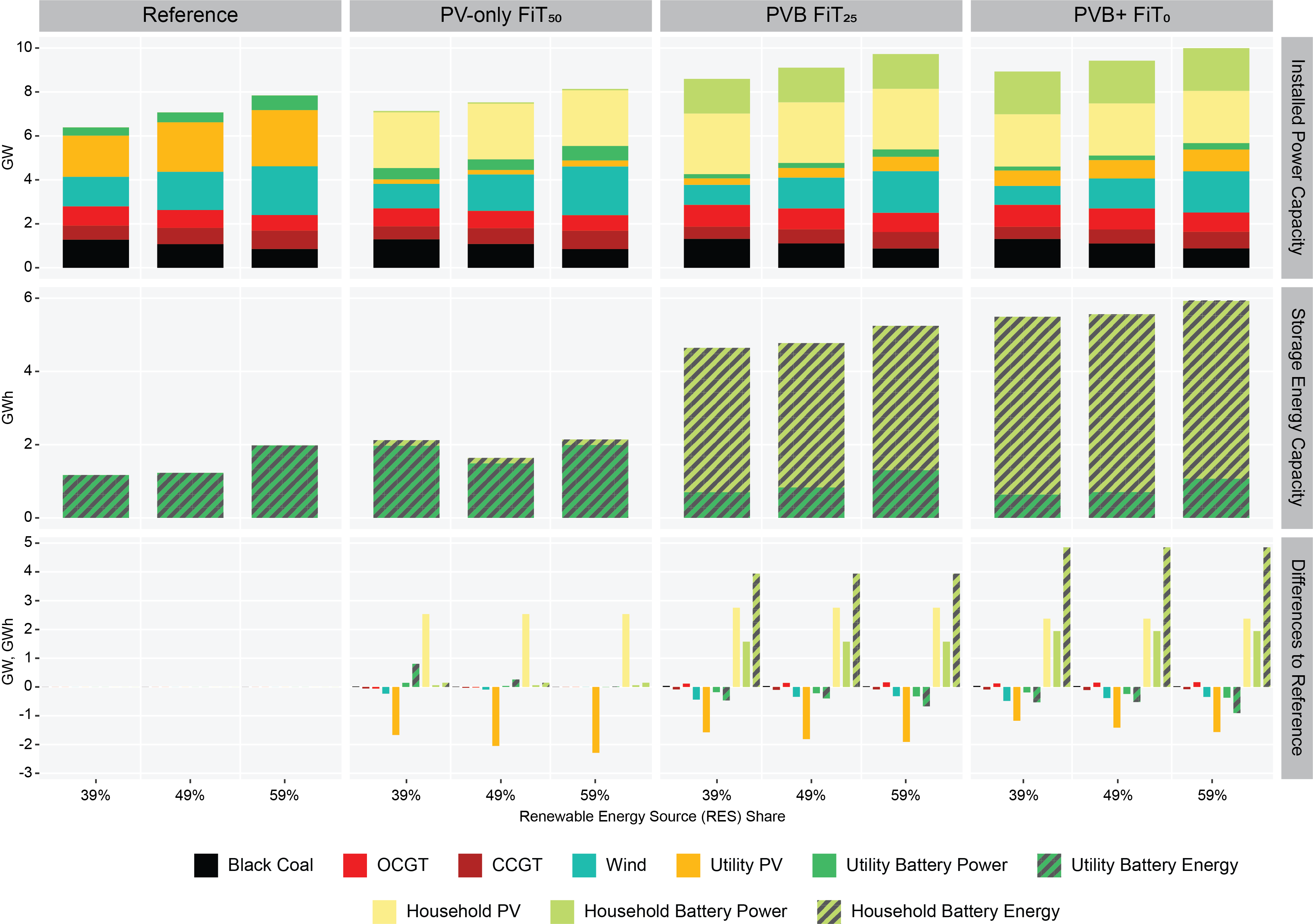}\\
\protect\caption{\label{fig: capacity changes batt20m} Installed power and storage energy capacity for varying FiT and RES shares (500,000 households) and the change in capacity with respect to the equivalent reference scenario, sensitivity with -20\% battery system costs.}
\end{figure}

\begin{figure}[h!]
\centering{} \includegraphics[width=0.82\textwidth]{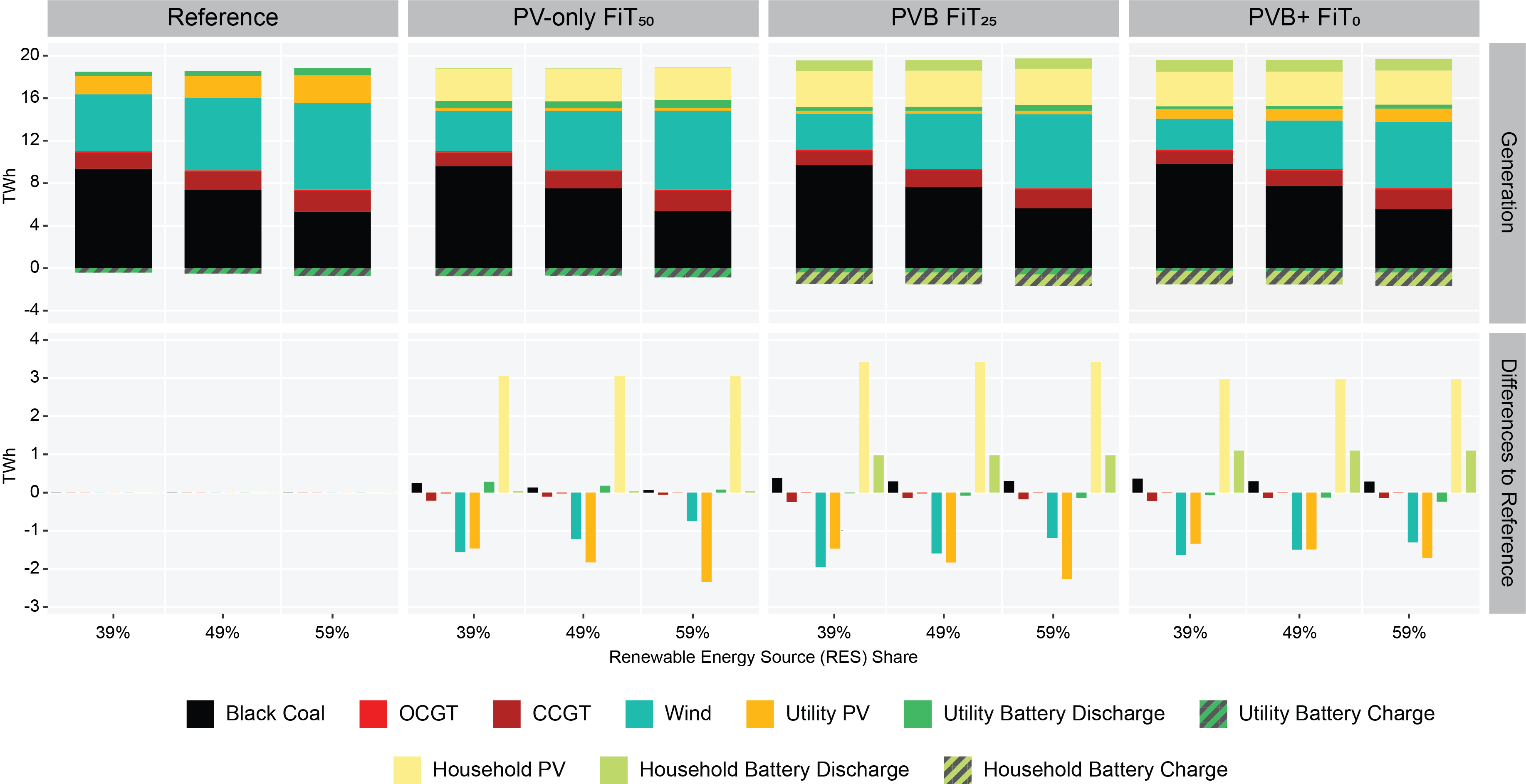}\\
\protect\caption{\label{fig: generation changes batt20m} Yearly generation for varying FiT and RES shares (500,000 households) and the change in generation with respect to the equivalent reference scenario, sensitivity with -20\% battery system costs.}
\end{figure}

A 20\% reduction in battery system costs leads to greater utility and household battery capacities. Household PV capacity also increases slightly to take advantage of the increased energy storage capacity (Figure \ref{fig: capacity changes batt20m}). Utility PV capacity remains heavily displaced by prosumage while wind continues to remain the dominant source of renewable energy generation (Figure \ref{fig: generation changes batt20m}). Qualitatively, the system implications from household PV battery prosumage remain consistent.

\pagebreak

\begin{figure}[h!]
\centering{} \includegraphics[width=0.82\textwidth]{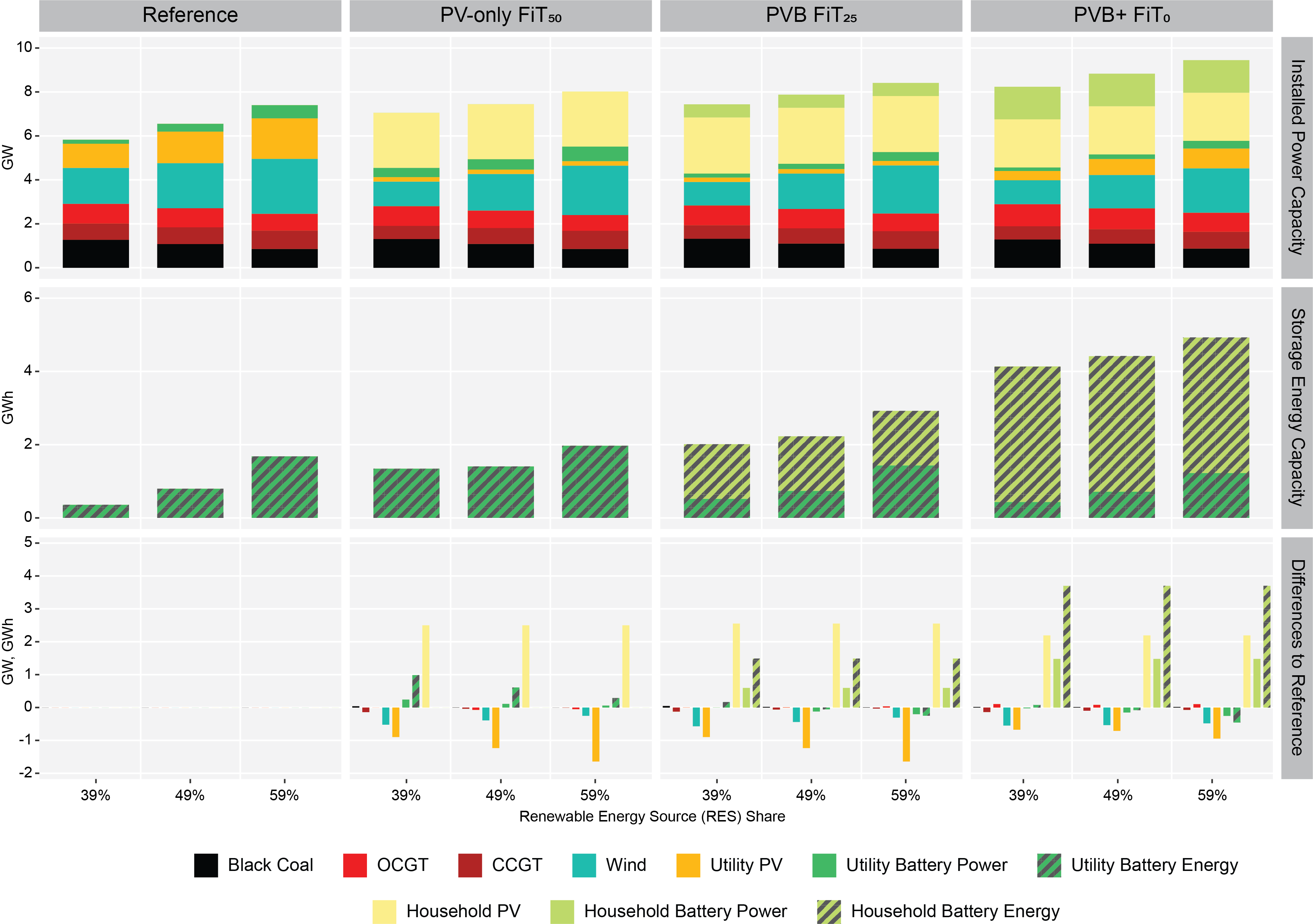}\\
\protect\caption{\label{fig: capacity changes batt20p} Installed power and storage energy capacity for varying FiT and RES shares (500,000 households) and the change in capacity with respect to the equivalent reference scenario, sensitivity with +20\% battery system costs.}
\end{figure}

\begin{figure}[h!]
\centering{} \includegraphics[width=0.82\textwidth]{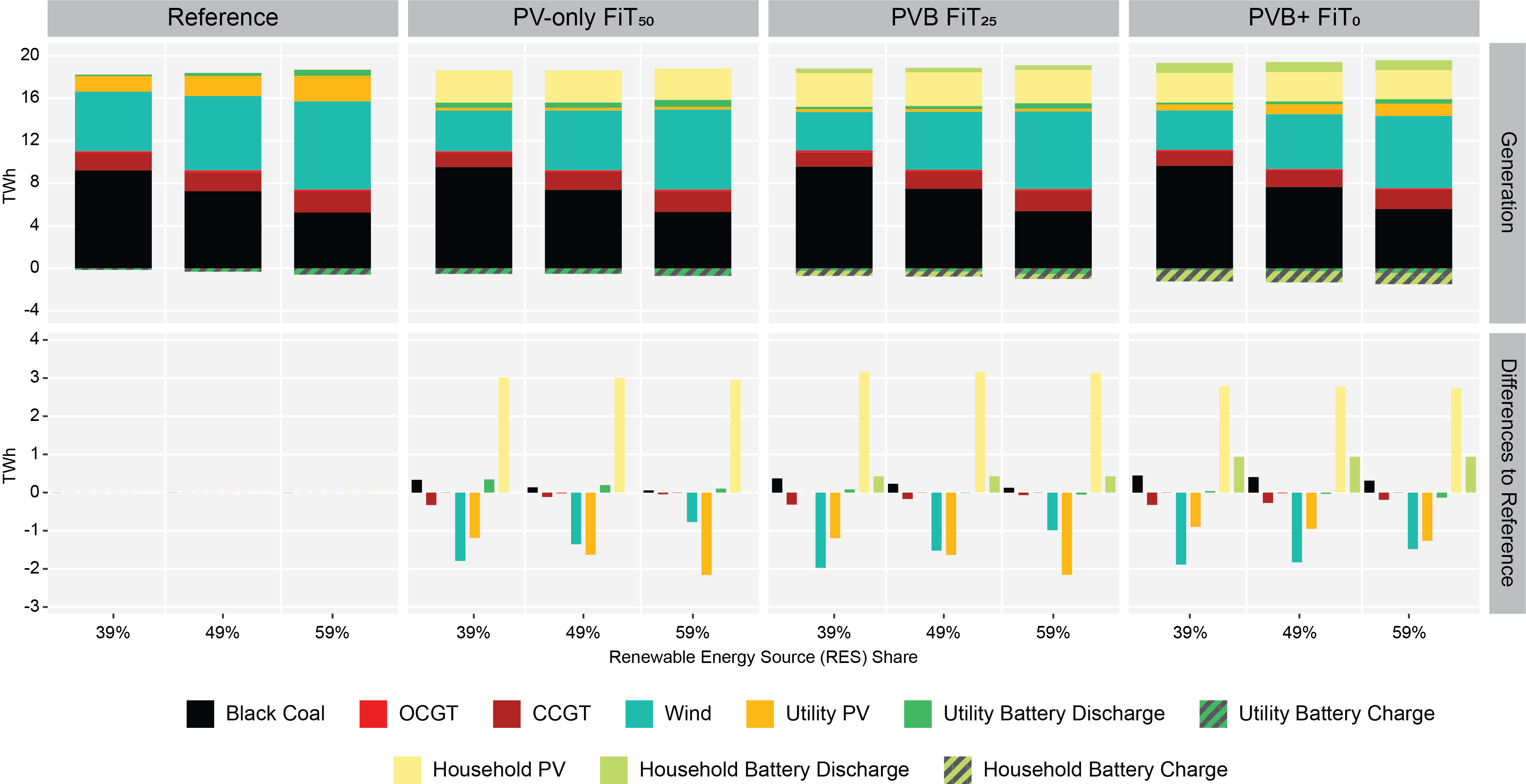}\\
\protect\caption{\label{fig: generation changes batt20p} Yearly generation for varying FiT and RES shares (500,000 households) and the change in generation with respect to the equivalent reference scenario, sensitivity with +20\% battery system costs.}
\end{figure}

A 20\% increase in battery system costs leads to reductions in utility and household battery capacities. Household PV capacity is also reduced (Figure \ref{fig: capacity changes batt20p}). Utility PV capacity remains heavily displaced while wind continues to remain the dominant source of renewable energy generation (Figure \ref{fig: generation changes batt20p}). Qualitatively, the system implications from household PV battery prosumage again remain consistent.

\pagebreak

\subsection{Number of prosumage households  ($\pm~20\%$)}

\begin{figure}[h!]
\centering{} \includegraphics[width=0.82\textwidth]{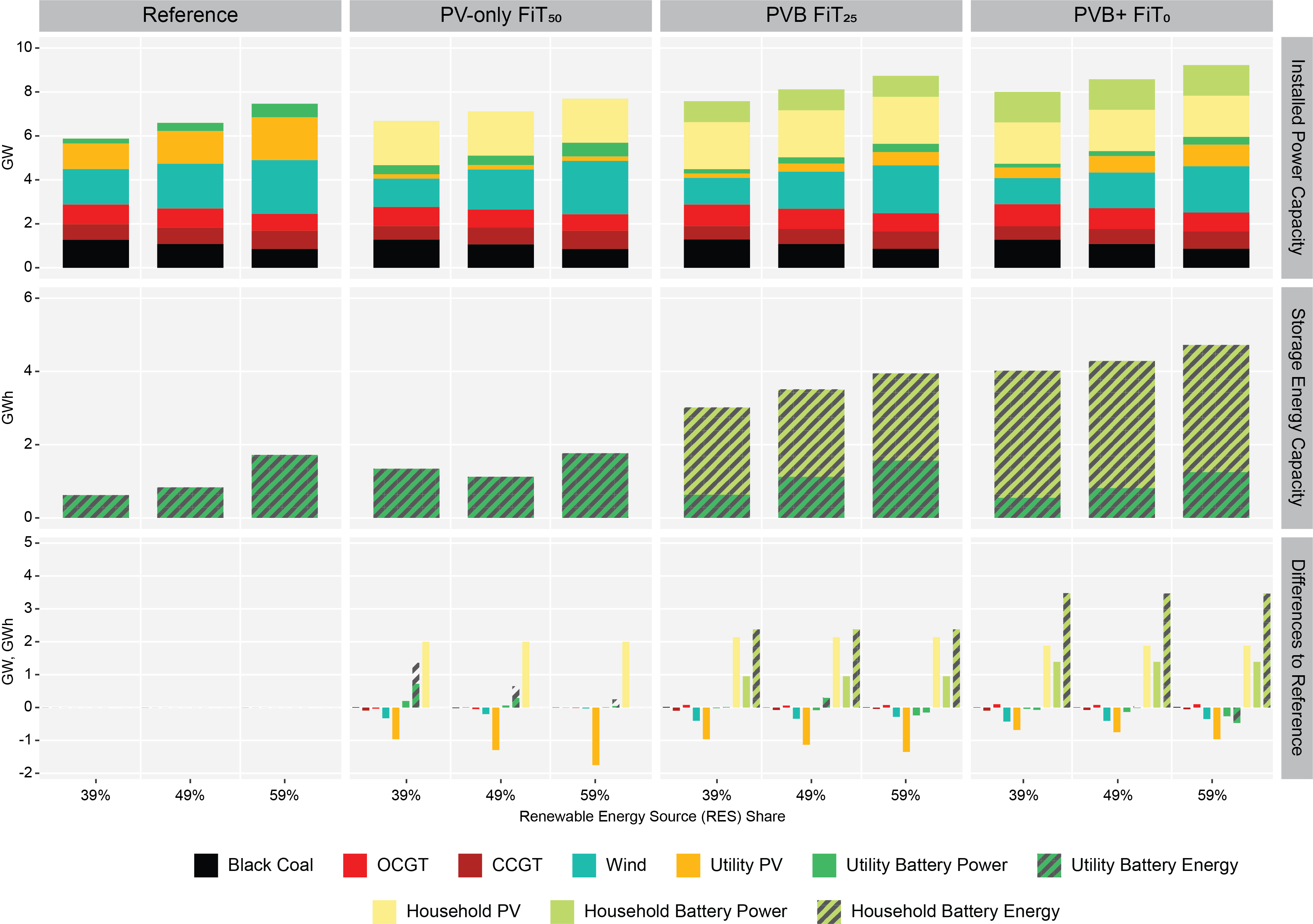}\\
\protect\caption{\label{fig: capacity changes 400000 hh} Installed power and storage energy capacity for varying FiT and RES shares (400,000 households) and the change in capacity with respect to the equivalent reference scenario.}
\end{figure}

\begin{figure}[h!]
\centering{} \includegraphics[width=0.82\textwidth]{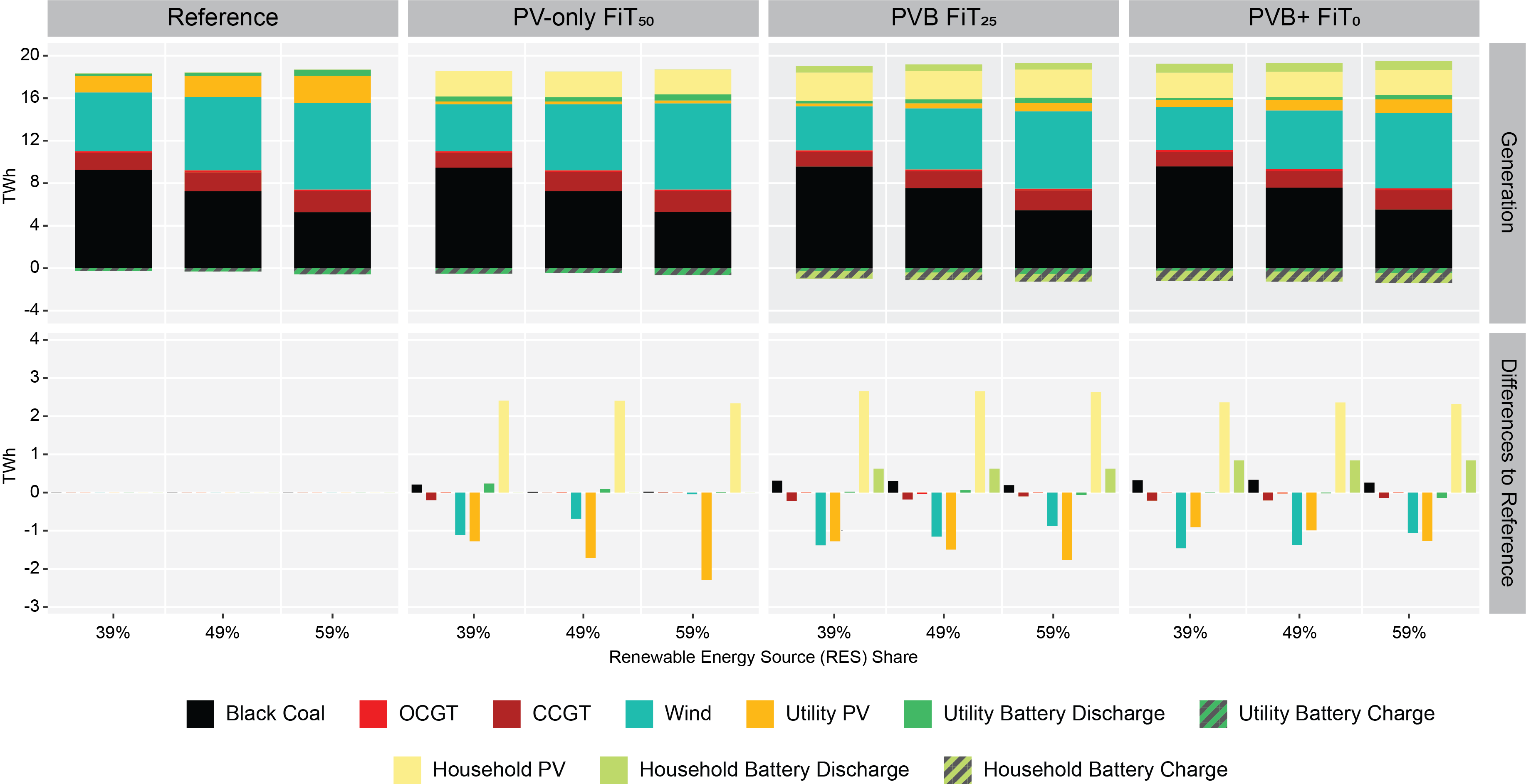}\\
\protect\caption{\label{fig: generation changes 400000 hh}Yearly generation for varying FiT and RES shares (400,000 households) and the change in generation with respect to the equivalent reference scenario.}
\end{figure}

Lowering the number of prosumage households from 500,000 to 400,000 reduces the displacement of wind and utility PV capacity and allows most of the utility PV capacity to recover in the PVB/PVB+ prosumage scenarios (Figure \ref{fig: capacity changes 400000 hh}). Wind generation continues to dominate as the primary source of renewable energy generation (Figure \ref{fig: generation changes 400000 hh}). Qualitatively, the system implications from household PV battery prosumage remain consistent.

\pagebreak

\begin{figure}[h!]
\centering{} \includegraphics[width=0.82\textwidth]{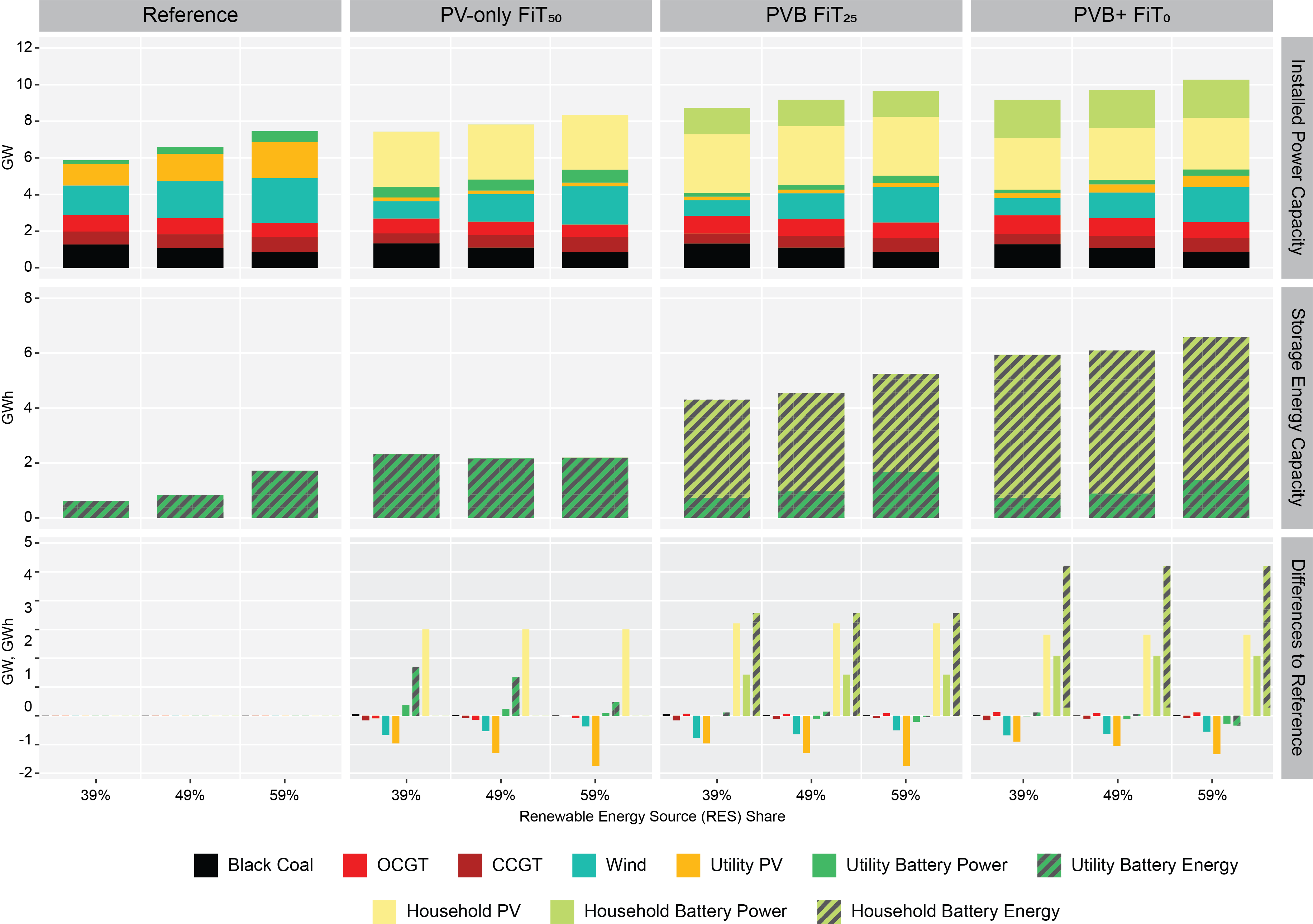}\\
\protect\caption{\label{fig: capacity changes 600000 hh}Installed power and storage energy capacity for varying FiT and RES shares (600,000 households) and the change in capacity with respect to the equivalent reference scenario.}
\end{figure}

\begin{figure}[h!]
\centering{} \includegraphics[width=0.82\textwidth]{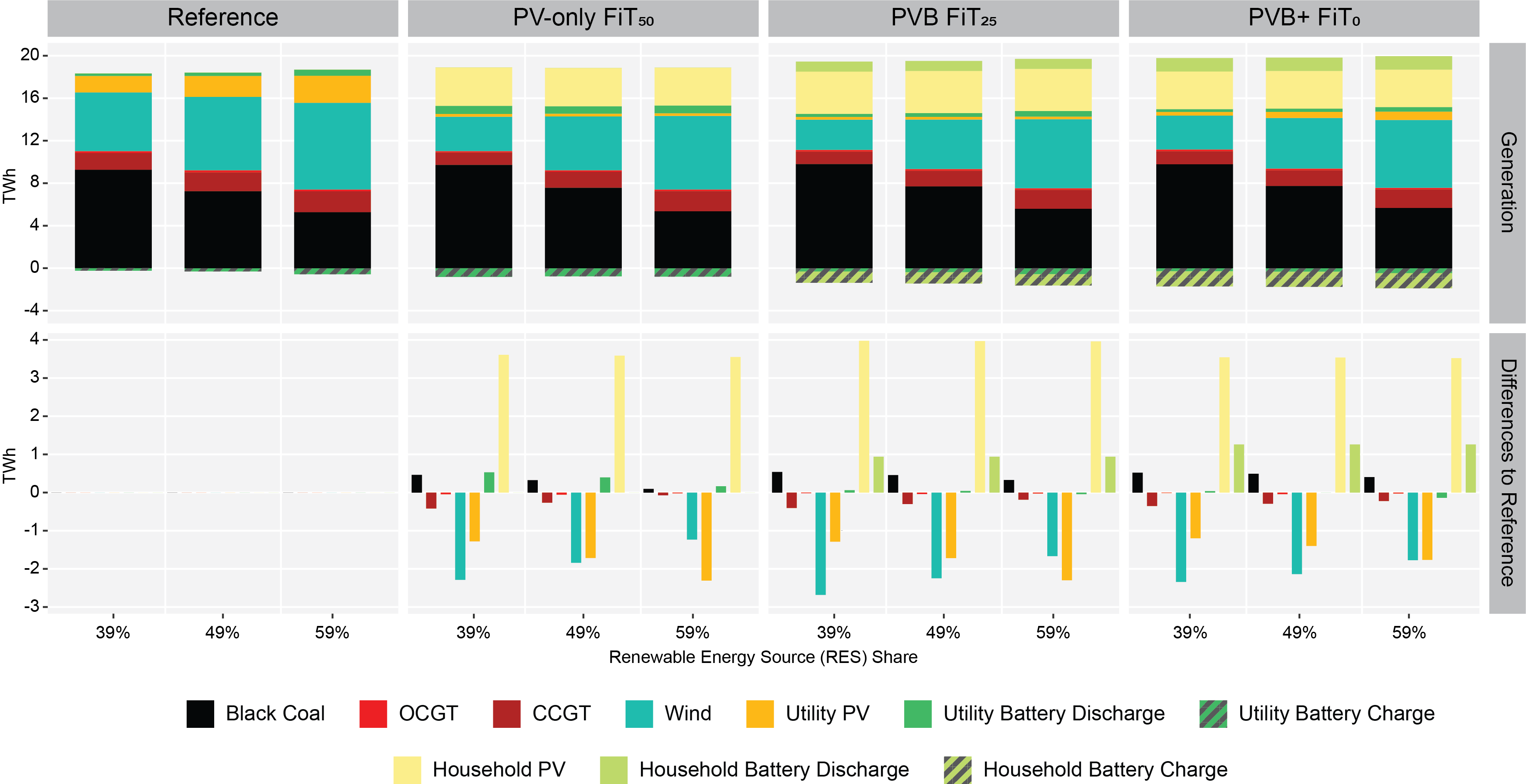}\\
\protect\caption{\label{fig: generation changes 600000 hh}Yearly generation for varying FiT and RES shares (600,000 households) and the change in generation with respect to the equivalent reference scenario.}
\end{figure}

Raising the number of prosumage households from 500,000 to 600,000 displaces further utility PV and wind capacity, while also slightly increasing utility battery capacity (Figure \ref{fig: capacity changes 600000 hh}). Wind generation continues to dominate as the primary source of renewable energy generation (Figure \ref{fig: generation changes 600000 hh}). Qualitatively, the system implications from household PV battery again prosumage remain consistent.

\end{document}